\documentclass{aa}  
\usepackage{longtable}
\pdfoutput=1
\usepackage{graphicx}
\usepackage{color}
\usepackage{txfonts}

\begin{document}

   \title{The ALMA-PILS survey: Complex nitriles towards IRAS 16293--2422}
   \titlerunning{Complex nitriles towards IRAS 16293--2422}

   \author{H. Calcutt\inst{1}, J. K. J{\o}rgensen\inst{1}, H. S. P. M\"uller\inst{2}, L. E. Kristensen\inst{1},  A. Coutens\inst{3}, T. L. Bourke\inst{4}, R. T. Garrod\inst{5}, M. V. Persson\inst{6}, M. H. D. van der Wiel\inst{7}, E. F. van Dishoeck\inst{8,9}, S. F. Wampfler\inst{10}}
\institute {\inst{1}Centre for Star and Planet Formation, Niels Bohr Institute \& Natural History Museum of Denmark, University of Copenhagen,
 {\O}ster \phantom{t}Voldgade 5--7, DK--1350 Copenhagen K., Denmark, Email: Calcutt@nbi.ku.dk\\
\inst{2}I. Physikalisches Institut, Universit\"at zu K\"oln, Z\"ulpicher Str. 77, 50937 K\"oln, Germany\\
\inst{3}Laboratoire d'Astrophysique de Bordeaux, Univ. Bordeaux, CNRS, B18N, all\'{e}e Geoffroy Saint-Hilaire, 33615 Pessac, France\\
\inst{4}SKA Organization, Jodrell Bank Observatory, Lower Withington, Macclesfield, Cheshire SK11 9DL, UK\\     
\inst{5}Departments of Chemistry and Astronomy, University of Virginia, Charlottesville, VA 22904, USA\\
\inst{6}Department of Space, Earth and Environment, Chalmers University of Technology, Onsala Space Observatory, 439 92, Onsala,
Sweden\\
\inst{7}ASTRON, the Netherlands Institute for Radio Astronomy, Postbus 2, 7990 AA Dwingeloo, The Netherlands\\
\inst{8}Leiden Observatory, Leiden University, PO Box 9513, 2300 RA Leiden, The Netherlands\\
\inst{9}Max-Planck Institut f\"{u}r Extraterrestrische Physik (MPE), Giessenbachstr. 1, 85748 Garching, Germany\\
\inst{10}Center for Space and Habitability (CSH), University of Bern, Sidlerstrasse 5, 3012 Bern, Switzerland\\
}
\authorrunning{H. Calcutt, J. K. J{\o}rgensen, H. S. P. M\"uller}

   \date{Received}
   % \abstract{}{}{}{}{} 
% 5 {} token are mandatory
 
  \abstract
   {Complex organic molecules are readily detected in the inner regions of the gaseous envelopes of forming protostars. Their detection is crucial to understanding the chemical evolution of the Universe and exploring the link between the early stages of star formation and the formation of Solar System bodies, where complex organic molecules have been found in abundance. In particular, molecules that contain nitrogen are interesting due to the role nitrogen plays in the development of life and the compact scales such molecules have been found to trace around forming protostars.}
   {The goal of this work is to determine the inventory of one family of nitrogen-bearing organic molecules, complex nitriles (molecules with a --C$\equiv$N functional group) towards two hot corino sources in the low-mass protostellar binary IRAS 16293--2422. This work explores the abundance differences between the two sources, the isotopic ratios, and the spatial extent derived from molecules containing the nitrile functional group.} 
   {Using data from the Protostellar Interferometric Line Survey (PILS) obtained with ALMA we determine abundances and excitation temperatures for the detected nitriles. We also present a new method for determining the spatial structure of sources with high line density and large velocity gradients --- Velocity-corrected INtegrated emission (VINE) maps. }
   {We detect methyl cyanide (CH$_3$CN) as well as 5 of its isotopologues, including the detection of CHD$_2$CN which is the first detection in the ISM. We also detect  ethyl cyanide (C$_2$H$_5$CN), vinyl cyanide (C$_2$H$_3$CN), and cyanoacetylene (HC$_3$N). We find that abundances are similar between IRAS 16293A and IRAS 16293B on small scales except for vinyl cyanide which is only detected towards the latter source. This suggests an important difference between the sources either in their evolutionary stage or warm-up timescales. We also detect a spatially double-peaked emission for the first time in molecular emission in the A source, suggesting that this source is showing structure related to a rotating toroid of material. }
   {With high-resolution observations we have been able to show for the first time a number of important similarities and differences in the nitrile chemistry in these objects. These illustrate the utility of nitriles as potential tracers of the physical conditions in star-forming regions.}

   \keywords{stars: formation -- stars: protostars -- ISM: Molecules -- ISM: individual objects: IRAS 16293$-$2422
               }

   \maketitle
%
%%
%________________________________________________________________
\section{Introduction}
Complex organic molecules have been detected in every environment associated with star formation. They are found at every stage from prestellar cores, to low-mass hot corinos (e.g. \citealt{jimenez-serra2016}) and high-mass hot cores (e.g. \citealt{Favre2017, Isokoski2013}), right through to protoplanetary disks (e.g. \citealt{Oberg2015,Walsh2016}). Their detection is critical to understanding the chemical development of star-forming regions and exploring the link between the early stages of star formation and the formation of Solar System bodies, where complex organic molecules have been found in abundance. 

Of particular interest to such studies is the detection of complex organic molecules containing nitrogen. They are significant due to the importance nitrogen plays in the development of life  \citep{Powner2009}, as well studies of fractionation, where the largest isotopic anomalies in the Solar System are found in highly volatile elements such as nitrogen (e.g. \citealt{Mumma2011, Wampfler2014}). The spectra of star-forming regions are rich in nitrogen-bearing molecules, such as methyl cyanide (CH$_3$CN), ethyl cyanide (C$_2$H$_5$CN) and isocyanic acid (HNCO). Methyl cyanide is considered a typical disk tracer in high-mass star formation \citep{Cesaroni2017}, owing to its excitation in very dense regions \citep{Beltran2005}. \\

In recent years it has become easier to take an inventory of the complex molecular content of star-forming regions at high angular and spectral resolution, providing an overview of the chemical evolution that occurs when a star forms. The Atacama Large Millimeter/submillimeter Array (ALMA) has been an indispensable tool in this effort, with its high sensitivity, angular resolution and spectral resolution capabilities, providing data which are orders of magnitude more sensitive than previous studies, on Solar System scales (e.g. \citealt{Jorgensen2016}). In this work we utilise ALMA observations from the Protostellar Interferometric Line Survey\footnote{\url{http://youngstars.nbi.dk/PILS/}} (PILS) of the protostellar binary IRAS 16293--2422 (hereafter IRAS 16293) to study nitriles. IRAS 16293 is a Class 0 solar-type protostar, which is often considered the best astrochemical testbed for low-mass star formation (see e.g. \citealt{vanDishoeck1995}; \citealt{Ceccarelli2000}; \citealt{Schoier2002}), due to its very bright line emission and relatively close location (120\,pc; \citealt{Loinard2008}), in the Ophiuchi star-forming complex. It consists of at least two protostars, A and B, separated by about 5\arcsec (600\,AU), which have been observed to be rich in complex organic molecules (\citealt{Bottinelli2004}; \citealt{Kuan2004}; \citealt{Bisschop2008}; \citealt{Jorgensen2011}). Continuum observations in ALMA bands 3, 6, and 7 show clear elongated emission towards IRAS 16293A in the direction of the reported velocity gradient in this source \citep{Pineda2012}, making it appear as a flattened edge-on rotating structure \citep{Jorgensen2016}. Source B conversely exhibits a near face-on structure evident in both its continuum emission and the narrow line widths observed in its spectra without clear velocity gradients indicative of rotation, with infall signatures observed towards the peak continuum position in several molecules \citep{Chandler2005,Pineda2012}.\\

Several nitrogen-bearing molecules have been studied previously in this source, with both single dish and interferometric observations. Interferometric observations of HNCO and CH$_3$CN emission were analysed by \citet{Bisschop2008} to explore excitation properties and spatial distributions. They found the molecular emission was compact towards both sources, with the bulk of emission originating from IRAS 16293A. They also detected CH$_3$$^{13}$CN in both sources to determine the $^{12}$C/$^{13}$C ratio. Severe line blending in the A source made determining accurate line-intensities difficult, however, blending in the B source was less severe resulting in a CH$_3$CN/CH$_3$$^{13}$CN ratio of $\sim$10. This low value compared to the local ISM value ($\sim$70; \citealt{Milam2005}) was attributed to CH$_3$CN emission being optically thick. 

 \citet{Jaber2017} also explored nitrogen-bearing molecules in IRAS 16293 using single-dish observations from the TIMASS survey \citep{Caux2011}. They took a census of the cyanopolyynes (species of the form HC$_n$N) in this source, to derive how their abundance varies across the protostellar envelope. This led to the first detection of DC$_3$N in a solar-type protostar. A number of first detections of nitrogen-bearing molecules have also occurred using the PILS dataset. \citet{Ligterink2017} detected methyl isocyanate (CH$_3$NCO) and \citet{Coutens2016} detected three singly-deuterated forms of formamide (NH$_2$CDO, cis-NHDCHO, trans-NHDCHO) and deuterated isocyanic acid (DNCO) for the first time in the ISM. Cyanamide (NH$_2$CN) was also detected for the first time in a solar-type protostar by \citet{Coutens2018}.

In this work, we focus on one family of nitrogen-bearing molecules, the nitriles. Nitriles are any organic compound that contains the --C$\equiv$N functional group. Their abundance in star-forming regions offers the opportunity to detect multiple nitrile molecules, highlighting important differences in the chemistry of these related molecules. Their detection is also important in understanding prebiotic chemistry. Simple nitriles, such as HCN, are key precursors to the prebiotic synthesis of RNA and DNA \citep{Banerjee2014}. More complex nitriles, such as vinyl cyanide (C$_2$H$_3$CN), have also been suggested as the building blocks of non-terrestrial cell membranes capable of forming and functioning in liquid methane at cryogenic temperatures \citep{Stevenson2015}, highlighting their significance in astrobiology as well as astrochemistry.

We present the most detailed study of nitriles in IRAS 16293 to date, exploring the chemistry in both sources A and B to explore the excitation, abundances and spatial scale of this family of organics. In Section \ref{sec:obs}  the observations used for the analysis are described, in Section \ref{sec:res} the analysis methodology and the results from this work are presented, and the significance of this work is discussed in Section \ref{sec:dis}. A summary and conclusions are presented in Section \ref{sec:con}.

\section{Observations} \label{sec:obs}
ALMA observations taken as part of the PILS program \citep{Jorgensen2016} are used for the analysis in this work.  The data are a combination of 15 hours of observations with the main array (typically 36--41 antennas in the array at the time of observations) and 30 hours of observations with the Atacama Compact Array (ACA; typically 8--10 antennas) in band 7, between 329.147\,GHz and 362.896\,GHz. The combined dataset analysed in this paper was produced with a circular restoring beam of 0\farcs5 at a spectral resolution of 0.2\,km\,s$^{-1}$  and an RMS of about  7--10\,mJy\,beam$^{-1}$\,channel$^{-1}$, i.e. approximately
4--5\,mJy\,beam$^{-1}$\,km\,s$^{-1}$. Further details of the data reduction and continuum subtraction procedure can be found in \citet{Jorgensen2016}.

\begin{figure*} 
\begin{center} 
\includegraphics[width=9cm, angle=0, clip =true, trim = 1cm 0cm 1cm 1cm]{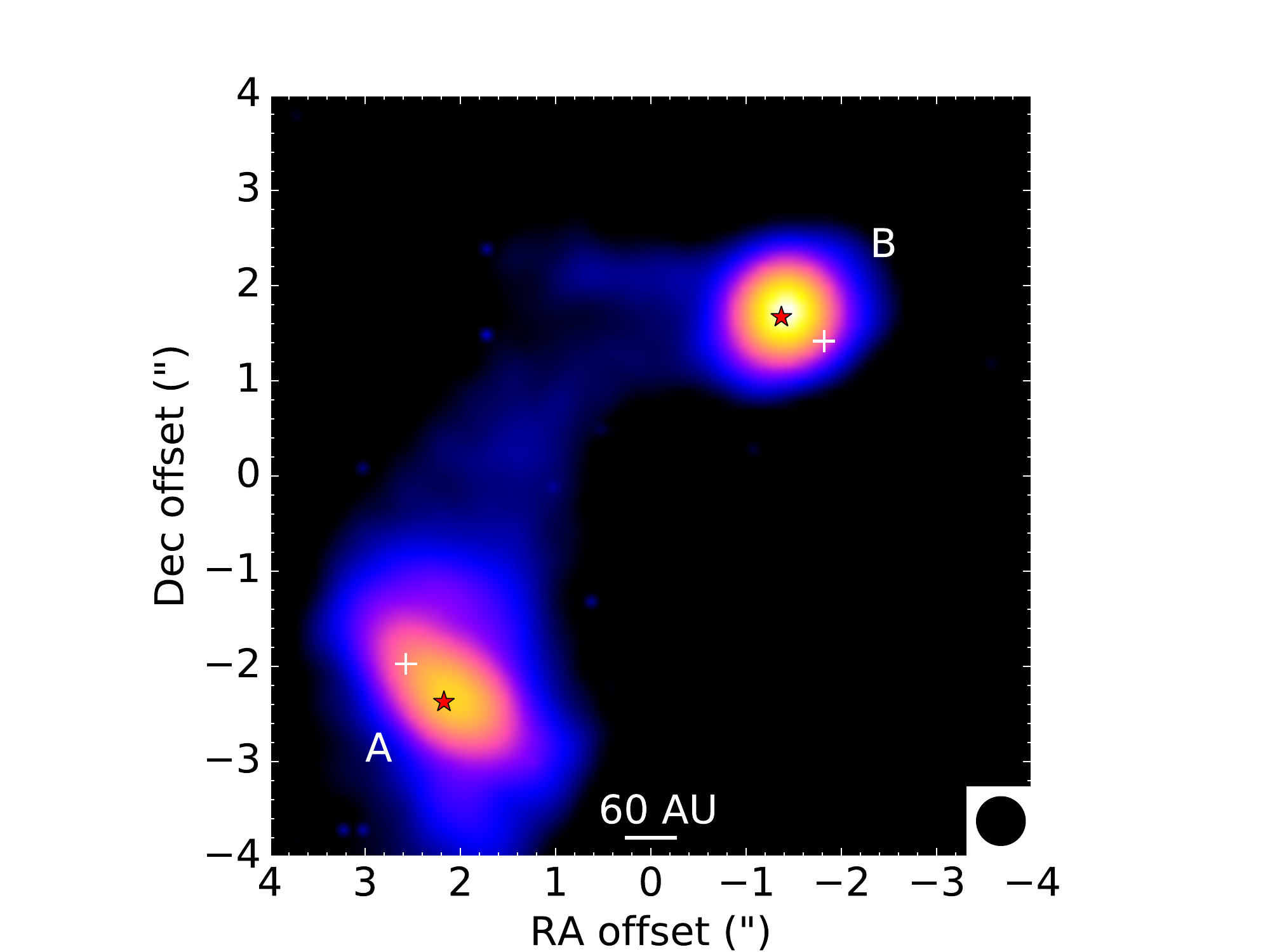}
\includegraphics[width=9cm, angle=0, clip =true, trim = 0cm 0cm 1cm 1cm]{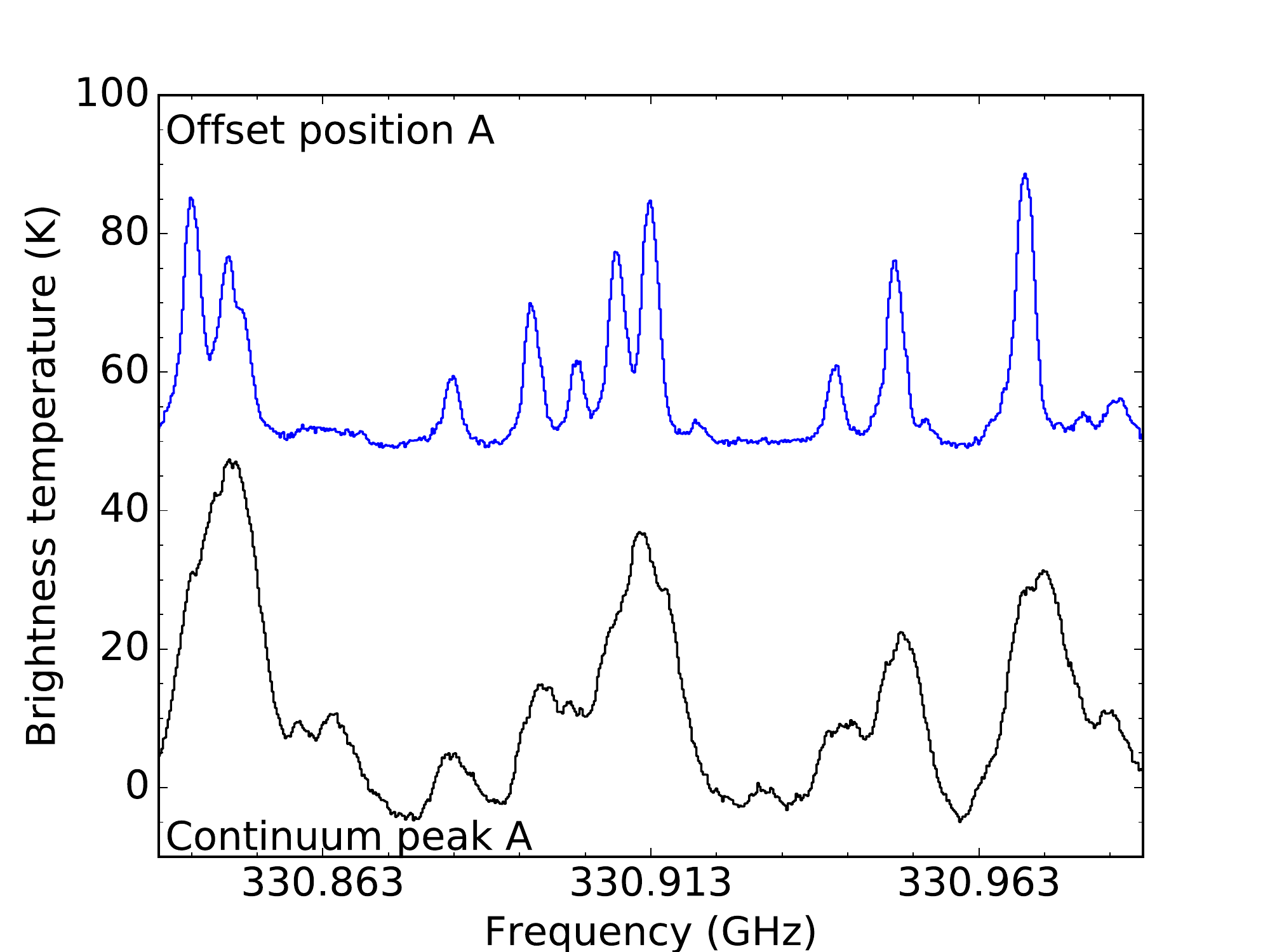}
 \end{center} 
 \caption{{\it Left panel:} Band 7 (0.87\,mm) continuum image of IRAS 16293 with the peak continuum positions marked with red stars and the offset positions analysed in this work marked with white crosses. {\it Right panel:} Spectra towards the continuum peak position of source A (black line) and the offset position (blue line) analysed in this work.\label{fig:sourceA}} 
 \end{figure*}

The spectral analysis presented in this work is performed towards two positions,  at $\alpha_{J2000}$=16$^{\rm h}$32$^{\rm m}$22{\rm \fs}58, $\delta_{J2000}$=$-$24$^{\circ}$28\arcmin32\farcs8 and $\alpha_{J2000}$=16$^{\rm h}$32$^{\rm m}$22{\rm \fs}90, $\delta_{J2000}$=$-$24$^{\circ}$28\arcmin36\farcs2. The first position is offset by one beam diameter (0\farcs5, 60\,AU)  from the continuum peak position of IRAS 16293B in the south west direction. This position is used as the lines are particularly bright, do not have strong absorption features towards the bright continuum of IRAS 16293B, and do not suffer from high continuum optical depth (\citealt{Coutens2016}, \citealt{Lykke2017}). The second position is offset by 0\farcs6 (68\,AU) north east from the peak continuum position. The $FWHM$ of lines around IRAS 16293A are much broader than around IRAS 16293B, particularly at the peak continuum emission position (3\,km\,s$^{-1}$ vs 1\,km\,s$^{-1}$). This makes line blending a significant problem for identification and analysis of molecular emission. At the IRAS 16293A offset position, however, the typical $FWHM$ is 2.2\,km\,s$^{-1}$, which  minimises line blending issues, but is still close enough to the peak position to detect weaker isotopologues. Both these positions are indicated with white cross in the left panel of Figure~\ref{fig:sourceA}, as well as the continuum peak position for both sources, which is indicated with a red star.  The right panel of Figure~\ref{fig:sourceA} shows the typical $FWHM$ of IRAS 16293A at the continuum peak position and the offset position.

\section{Results and analysis} \label{sec:res}
The spectra  extracted from both sources A and B have been searched for complex nitrogen-bearing organics. To determine abundances and excitation temperatures, the spectral modelling software CASSIS\footnote{\url{http://cassis.irap.omp.eu/}} has been used to compute the best fit spectral models to the observations, assuming local thermodynamic equilibrium (LTE). Line lists for each of the molecules discussed in this work were taken from the CDMS catalogue\footnote{\url{http://www.astro.uni-koeln.de/cdms}}. The CH$_3^{13}$CN, $^{13}$CH$_3$CN and CH$_3$C$^{15}$N entries are based on \citet{MeCN_rot_isos_2009} with additional data in the range of our survey from \citet{MeCN_isos_rot_1996}. The CH$_2$DCN and CHD$_2$CN entries are based on \citet{CH2DCN_CHD2CN_rot_2013} with additional data from \citet{CH2DCN_rot_1992} and \citet{MeCN_rot_isos_2009} for CH$_2$DCN and from \citet{CHD2CN_CHD2NC_rot_1978} for CHD$_2$CN. Vibrational corrections were taken from the main isotopic species \citep{MeCN_v8le2_2015}; differences among the isotopic species can be neglected at the excitation temperatures of the species presented in this work.\\

The ethyl cyanide entry is based on \citet{EtCN_rot_2009} with additional important contributions from \citet{EtCN_rot_1996} and from \citet{EtCN_rot_1994}. Vibrational correction factors to the partition function are provided in the CDMS; they 
are 1.32 and 1.46 for 140 and 160\,K, respectively \citep{EtCN_IR_1981}. The cyanoacetylene entry is based on \citet{HC3N_vibs_rot_2000} with important additional contributions, also in the range of our survey, from \citet{HC3N_rot_1995}. Vibrational correction factors are 1.09 and 1.38 at 100 and 160\,K, respectively \citep{HC3N_DC3N_IR_1976}. The vinyl cyanide entry was taken from the CDMS. It is based on \citet{VyCN_isos_rot_2008} 
with additional data in the range of our survey from \citet{VyCN_rot_1994}. Numerous low-lying 
excited vibrational states were taken into account in the calculation of the partition function 
(\citealt{VyCN_IR_1999} and H.~S.~P. M{\"u}ller, unpublished). The correction factor is 1.17 at 150\,K. \\

A large grid of models was run to determine the best fit excitation temperature, $FWHM$, velocity peak $V\rm{_{peak}}$, and column density for each molecule. Excitation temperatures of 90 -- 300\,K are explored for both sources. A $FWHM$ of between 1.8 -- 2.4\,km\,s$^{-1}$ for the A source and 0.8 -- 1.2\,km\,s$^{-1}$ for the B source is explored.  A $V\rm{_{peak}}$ of between 0.6 -- 0.8\,km\,s$^{-1}$ for the A source, based on the observed velocity shift at the offset position for source A of 0.8\,km\,s$^{-1}$ is explored. For the B source a $V\rm{_{peak}}$ of between 2.5 -- 2.9\,km\,s$^{-1}$ is explored, based on the $V\rm{_{LSR}}$ of 2.7\,km\,s$^{-1}$ for source B. Column densities in the range of 1$\times$10$^{13}$ -- 1$\times$10$^{17}$ cm$^{-2}$ are explored. The reduced $\chi^2$ minimum is then computed to determine the best fit model, assuming a source size of 0\farcs5 determined from emission maps (see Figure \ref{fig:map}). The significant spectral density seen in IRAS 16293 means that line blending is an issue within the spectra for both sources A and B. Significantly blended lines, i.e. lines which fail the Rayleigh criterion, where the minimum distance between two spectral lines must be such that the maximum intensity of one line falls on the first minimum of the other line \citep{Snyder2005}, are excluded from the best fit model calculations. Optically thick lines are also a problem for some of the species modelled, and therefore only optically thin lines ($\tau$<0.2) are used to determine the best fit model. \\

Due to the strong continuum reported towards IRAS 16293, the continuum component is included in the spectral model as a background temperature component. Since the continuum is dominated by the high column density dust in the disk while the lines arise from the infalling gas in the envelope, the dust and gas are not fully coupled at the positions analysed in this work. The correction is calculated as in \citet{Jorgensen2016} and is 1.14 at 125\,K towards the B position and 1.24 at 125\,K towards the A position.\\

The main sources of error for the column density and excitation temperatures are from the quality of the fit of the spectral model to the data, which relies on the assumption of LTE, optically thin emission and Gaussian profiles. This is particularly a problem for species which have a number of optically thick lines which had to be excluded from the fit, and therefore a lower number of lines to constrain the fitted parameters. We estimate the errors, based on varying the $N_{\rm tot}$ and $T_{\rm ex}$ to determine the impact this has on the fit of the observations.

\subsection{Column densities and excitation temperatures}		

Transitions of several prominent complex nitrogen-bearing organics are detected towards both IRAS 16293A and IRAS 16293B. Table \ref{tab:coldens} shows the best-fit excitation temperatures, column densities and relative abundance for each species in each source. Lines towards IRAS 16293A are best fitted with a $FWHM$ of 2.2 km\,s$^{-1}$ and a $V\rm{_{peak}}$ of 0.8 km\,s$^{-1}$ except for HC$_3$N which is best fitted with a $V\rm{_{peak}}$ of 1.2 km\,s$^{-1}$. Lines towards IRAS 16293B are best fitted with a $FWHM$ of 1.0 km\,s$^{-1}$ and a $V\rm{_{peak}}$ of 2.7 km\,s$^{-1}$. The transitional information for every detection presented in this work can be found in the tables in Appendix \ref{sec:trans}.

\begin{table*}

\caption{Excitation temperature ($T_{\rm ex}$), column density ($N_{\rm tot}$) and relative abundance determined relative to CH$_3$CN for complex nitrogen bearing species in IRAS 16293B and IRAS 16293A.}\label{tab:coldens}
\centering
\begin{tabular}{cccc}
\hline
\hline
Molecule&$T_{\rm ex}$&$N_{\rm tot}$  &Relative abundance\\
&(K)&(cm$^{-2}$)\\
\hline
&&IRAS 16293A&\\
\hline
CH$_3$CN$^{\dagger}$&120 (10)&8.0 (1.0)$\times$10$^{16}$&1\\
$^{13}$CH$_3$CN&120 (10)&1.0 (0.2)$\times$10$^{15}$&0.013\\
CH$_3$$^{13}$CN&120 (10)&1.2 (0.2)$\times$10$^{15}$&0.015\\
CH$_3$C$^{15}$N&120 (10)&3.4 (0.2)$\times$10$^{14}$&0.004\\
CH$_2$DCN&120 (10)&3.5 (0.5)$\times$10$^{15}$&0.044\\
CHD$_2$CN&120 (10)&2.2 (0.5)$\times$10$^{14}$&0.003\\
C$_2$H$_5$CN&140 (10)&9.0 (0.2)$\times$10$^{15}$&0.113\\
C$_2$H$_3$CN&150 (10)&$\leq$2.4$\times$10$^{14\dagger\dagger}
$& \multicolumn{1}{l}{\phantom{111.1}$\leq$0.002$^{\dagger\dagger}$}\\
HC$_3$N&160 (20)&4.4 (0.2)$\times$10$^{14}$&0.006\\

\hline
&&IRAS 16293B&\\
\hline
CH$_3$CN$^{\dagger}$&110 (10)&4.0 (1.0)$\times$10$^{16}$&1\\
$^{13}$CH$_3$CN&110 (10)&6.0 (0.2)$\times$10$^{14}$&0.015\\
CH$_3$$^{13}$CN&110 (10)&5.0 (0.2)$\times$10$^{14}$&0.013\\
CH$_3$C$^{15}$N&110 (10)&1.6 (0.2)$\times$10$^{14}$&0.004\\
CH$_2$DCN&110 (10)&1.4 (0.2)$\times$10$^{15}$&0.035\\
CHD$_2$CN&110 (10)&2.0 (0.2)$\times$10$^{14}$&0.005\\
C$_2$H$_5$CN&110 (10)&3.6 (0.2)$\times$10$^{15}$&0.090\\
C$_2$H$_3$CN&110 (10)&7.4 (0.2)$\times$10$^{14}$&0.019\\
HC$_3$N&100 (20)&1.8 (0.2)$\times$10$^{14}$&0.005\\

\hline
\end{tabular}
\tablefoot{All models assume LTE and a source size of 0\farcs5. Errors are given in brackets. $^{\dagger}$Derived from the v$_8$=1 state. $^{\dagger\dagger}$Determined for the 3$\sigma$ limit.}
\end{table*} 

\begin{figure*} 
\begin{center} 
\includegraphics[width=12cm, angle=0, clip =true, trim = 5cm 5.2cm 4cm 4cm]{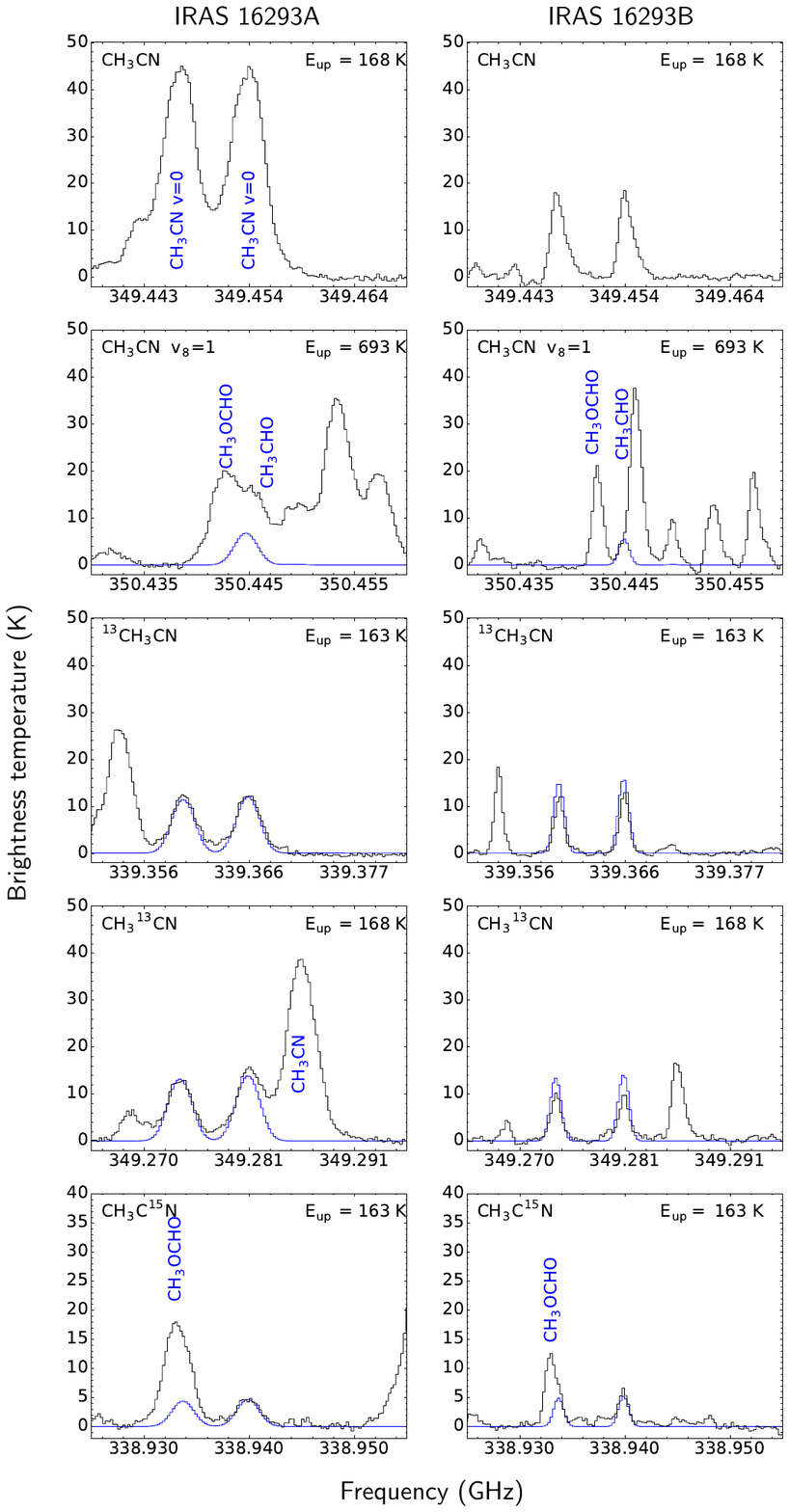}
\end{center} 
\caption{The spectra of the 19$_0$\,--\,18$_0$ line of CH$_3$CN, CH$_3$CN v$_8$=1, $^{13}$CH$_3$CN, CH$_3$$^{13}$CN, and CH$_3$C$^{15}$N in sources A and B. An LTE model for each species is overlaid in blue except for CH$_3$CN where the emission is optically thick and therefore a good fit cannot be found. Blended species are labelled on the figure. \label{fig:ch3cn}} 
\end{figure*}

Methyl cyanide (CH$_3$CN) is the most abundant complex nitrile detected in IRAS 16293, being at least one order of magnitude more abundant than the next most abundant cyanide, and found to be optically thick in both sources. Several transitions of the main species, both the v=0 state and v=1 excited state are detected towards both sources, as well as several of its isotopologues:  CH$_3$C$^{15}$N, $^{13}$CH$_3$CN, CH$_3$$^{13}$CN, CH$_2$DCN, and CHD$_2$CN (Figure \ref{fig:ch3cn}). The detection of the doubly deuterated form (CHD$_2$CN) is the first detection in the ISM. To compute the excitation temperature for these species we fit the most optically thin species (CH$_3$C$^{15}$N and CHD$_2$CN) allowing the excitation temperature to vary in the range 90\,--\,300 K. We then performed a second fit for all methyl cyanide species to determine their column densities with a fixed excitation temperature of the average determined from the first fit. The best fit models for methyl cyanide and its isotopologues are overlaid on the spectra for the 9 brightest lines in Figures \ref{fig:ch3cnv8=1}\,--\,\ref{fig:} in the Appendix. $^{13}$CH$_3$$^{13}$CN is not detected towards IRAS 16293A or B, however, its transitions within the observed frequency range are weak and could be obscured by line blending. The column densities of the two $^{13}$C forms of methyl cyanide are similar.
 
Vinyl cyanide is only detected in IRAS 16293B but an upper limit column density is determined towards source A, using $1.05{\times}RMS{\times}\sqrt{{\Delta}V{\times}FWHM}$ to compute the 1$\sigma$ limit, where ${\Delta}V$ is the spectral resolution of the data and 1.05 is a factor to account for a 5\% flux calibration uncertainty, and the $FWHM$ used is the same as other nitriles fitted at this position (2.2\,km\,s$^{-1}$). An upper limit column density is then determined using a 3$\sigma$ limit for the 331.087\,GHz line of vinyl cyanide (a bright line in a relatively line free region).

Ethyl cyanide and cyanoacetylene (HC$_3$N) are detected in both sources. Producing a good LTE model for HC$_3$N was complicated by there only being three lines available in the data to model, two of which (345.609\,GHz and 354.697\,GHz) are blended with other molecules (Figure \ref{fig:hc3n}). Such blending, which is more severe towards the A source, may explain why the V$_{peak}$ determined for HC$_3$N in the A source is 2 channels shifted compared to the other nitrile species, and numerous molecules from other PILS studies (\citealt{Ligterink2017}, Manigand et al. in prep). Alternatively, HC$_3$N could be partially tracing a different physical component as well as the hot corino. It is difficult to explore this further without more lines observed at high angular resolution. The column densities and excitation temperatures given in Table \ref{tab:coldens} are therefore only a best estimate. The isotopologues of these species were also searched for in both sources. Lines of C$_2$H$_5$$^{13}$CN can be fitted to features in the spectra of IRAS 16293B, however, these features are very weak and in all cases either severely blended or very near the noise level, and hence cannot be claimed as a detection. Isotopologues of vinyl cyanide and additional isotopologues of ethyl cyanide are not detected in this work. A search for HC$_5$N and DC$_3$N were also undertaken as these species have been identified in IRAS 16293 by \citet{Jaber2017}. Their transitions in the PILS dataset, however, are too weak to be detected.

\begin{figure*} 
\begin{center} 
\includegraphics[width=19cm, angle=0, clip =true, trim = 1.6cm 14cm 2cm 0cm]{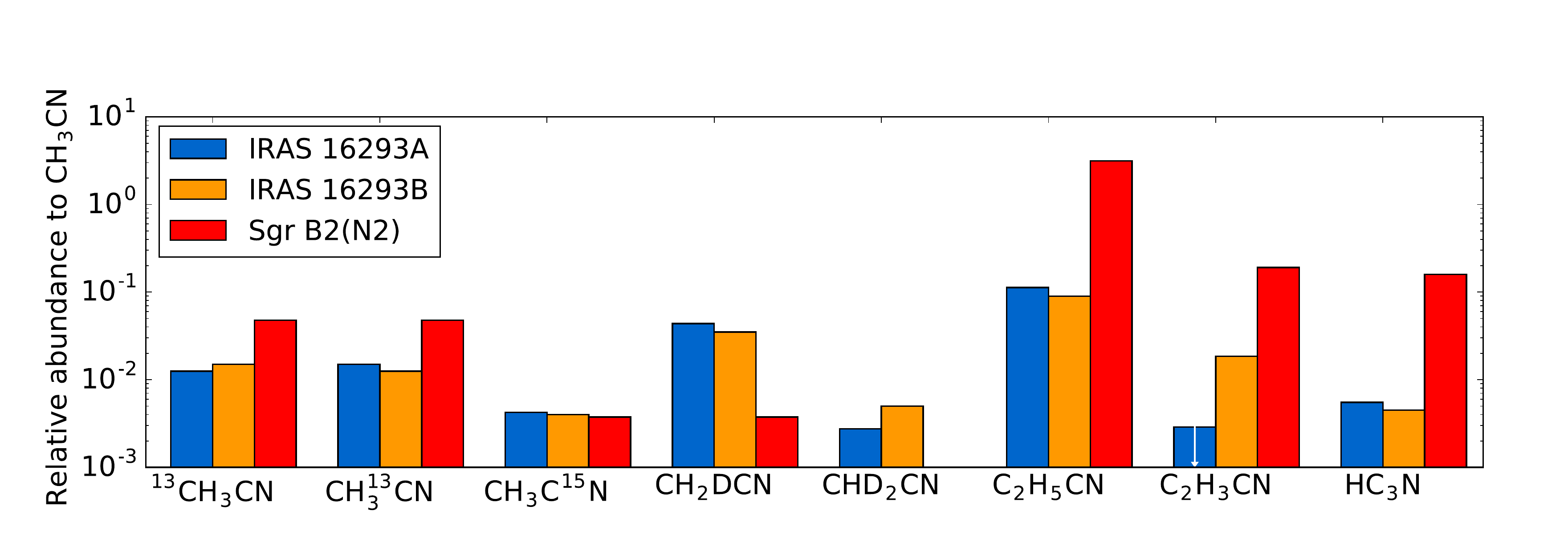}
\includegraphics[width=19cm, angle=0, clip =true, trim = 2cm 14cm 2cm 0cm]{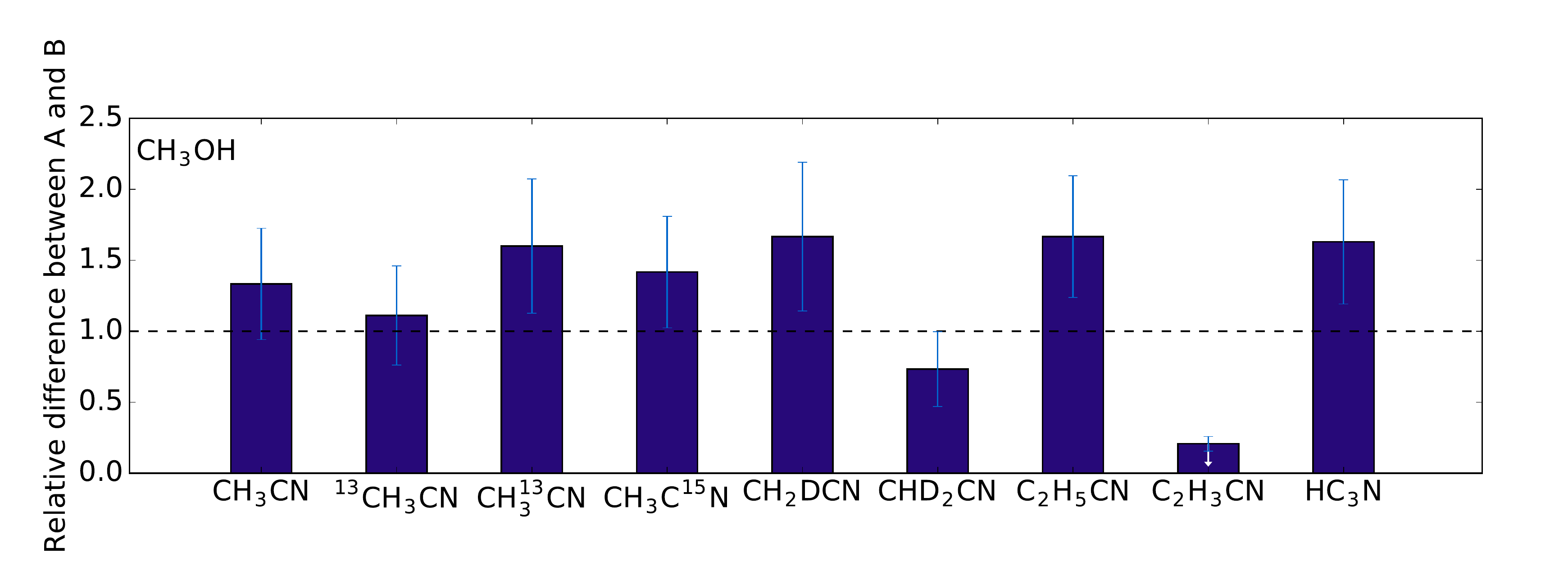}
\includegraphics[width=19cm, angle=0, clip =true, trim = 2cm 14cm 2cm 0cm]{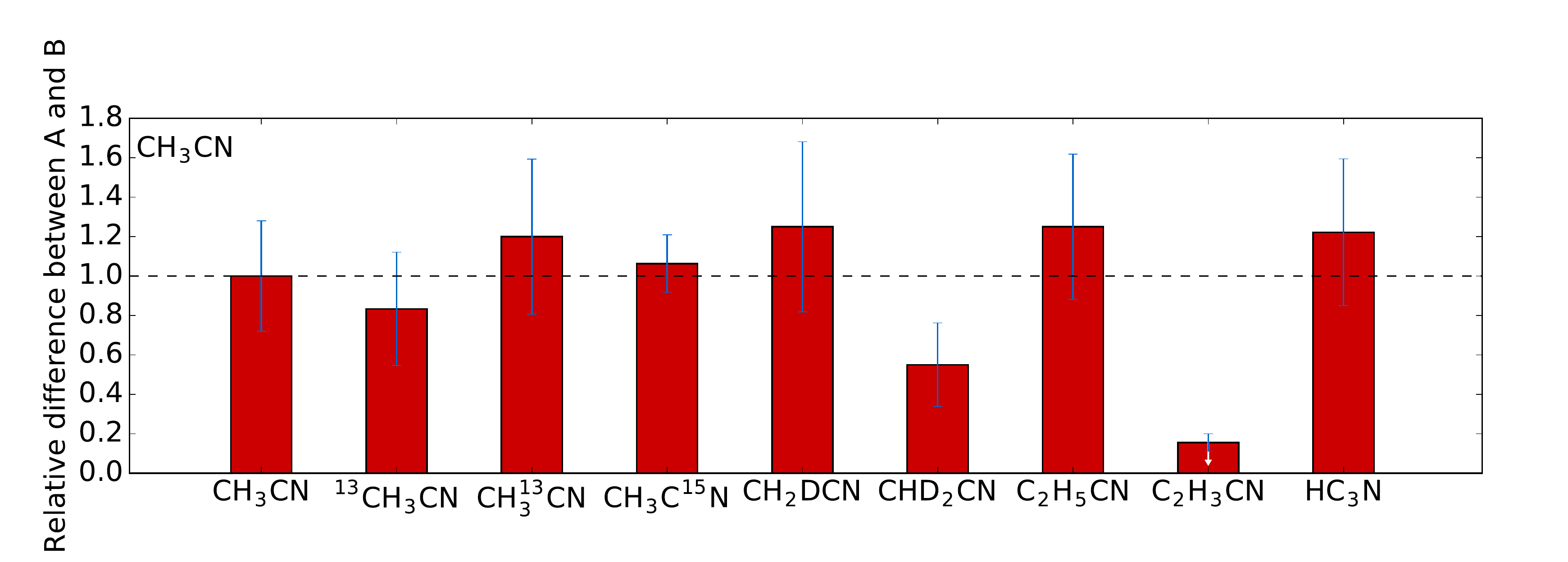}
 \end{center} 
 \caption{Top panel: The abundance of each species discussed in this work relative to methyl cyanide in IRAS 16293A and IRAS 16293B. Middle panel: Relative abundance difference between A and B (X(A)/X(B)) when computed relative to methanol. Bottom panel: Relative abundance difference between A and B (X(A)/X(B)) when computed relative to methyl cyanide.\label{fig:rel_abun}} 
 \end{figure*} 
\subsection{Relative abundances}\label{sec:rel}

To compare the column densities presented in the previous section for each source, the abundance of each molecule is computed relative to methyl cyanide and methanol. Since the transitions of CH$_3$CN v=0 are optically thick, the total column density for methyl cyanide is derived from the column density for the excited (v$_8$=1) state. For methanol the column densities are taken from \citet{Jorgensen2016} and J{\o}rgensen et al. (subm.) for IRAS 16293B and Manigand et al. (in prep.) for IRAS 16293A, which were both determined using column densities for CH$_{3}^{18}$OH multiplied by the standard $^{16}$O/$^{18}$O ISM value of 560 \citep{Wilson1994}.

In the top panel of Figure \ref{fig:rel_abun} the abundance of each species is plotted relative to methyl cyanide for both IRAS 16293A and IRAS 16293B. In the middle and bottom panels, the relative abundance ratio between A and B are plotted with respect to methanol abundances and methyl cyanide abundances, respectively.

For methyl cyanide and its isotopologues, the relative abundances are remarkably similar between A and B with differences in the abundance ratio in the range 0.6\,--\,1.7 (see Section \ref{sec:dis} for discussion). Abundances computed relative to methanol show more variation between the sources. The  difference in abundance ratio between A and B for ethyl cyanide and HC$_3$N is also similar to the difference relative to methyl cyanide.

Vinyl cyanide is the distinct outlier among the nitriles, being at least 9 times more abundant towards B, based on an upper limit column density calculated for the 331.087\,GHz line towards source A at an excitation temperature of 150\,K. Upper limit column densities were also computed for a range of excitation temperatures (50--500\,K) to determine whether this striking abundance difference could be due to differences in excitation. Whilst excitation temperatures of 280\,K did produce higher upper limit column densities for vinyl cyanide, plotting these column densities with CASSIS led to a clear overproduction of emission for some lines. We are therefore confident that the abundance differences for vinyl cyanide determined between IRAS 16293A and B are due to chemical differences and not excitation effects. 

\subsection{Isotopic ratios}\label{sec:isotopic}
 
The detection of several isotopologues of methyl cyanide in this work enables the computation of some isotopic ratios from the data. Since the main species of methyl cyanide is optically thick, the column density of the excited (v$_8$=1) state of methyl cyanide is used to compute the $^{12}$C/$^{13}$C, the $^{14}$N/$^{15}$N, and the D/H ratios in sources A and B in Table \ref{tab:iso}. \\
\begin{table}
\caption{Isotopic ratios in IRAS 16293A and B using methyl cyanide and its isotopologues}\label{tab:iso}
\centering
\begin{tabular}{cccccc}
\hline
\hline
Isotopic ratio&Source A& Source B\\
\hline
$^{12}$C/$^{13}$C$^{a}$&80$\pm$19&67$\pm$17 \\
$^{12}$C/$^{13}$C$^{a}$&67$\pm$14&80$\pm$20\\
$^{14}$N/$^{15}$N&235$\pm$33&250$\pm$70\\
D/H(D1,D0)$^{b}$&1.5$\pm$0.3\%&1.2$\pm$0.3\%\\
D/H(D2,D0)$^{b}$&3.0$\pm$0.3\%&4.0$\pm$0.5\%\\
\hline
\end{tabular}
\tablefoot{$^{a}$The $^{12}$C/$^{13}$C ratio is derived from the $^{13}$CH$_3$CN/CH$_3$CN and CH$_3^{13}$CN/CH$_3$CN ratios, respectively. $^{b}$The D/H ratio was derived from the CH$_2$DCN/CH$_3$CN and CHD$_2$CN/CH$_3$CN ratios, respectively, taking into account the statistical correction. With D1 referring to the singly-deuterated form, D2 referring to the doubly-deuterated form, and D0 referring to the main species.}
\end{table}

The $^{12}$C/$^{13}$C ratio is determined for both the $^{13}$CH$_3$CN and the CH$_3$$^{13}$CN isotopologues of methyl cyanide. The values towards both sources are similar to the ISM value of 68$\pm15$ (\citealt{Milam2005}). The $^{14}$N/$^{15}$N ratio is similar towards IRAS 16293A and B to values determined from HNC isotopologues by \citet{Wampfler2014} towards IRAS 16293A but higher than the value determined in the same work for HCN isotopologues, and the lower limits found for formamide and HNCO of 100 and 138 respectively by \citet{Coutens2016}. The D/H ratio given in the table is calculated using both the column densities of CH$_2$DCN and CHD$_2$CN, where both values have been corrected for the number of hydrogen to deuterium atoms.  The values range from 1.2--1.5\% for both sources when using singly-deuterated column densities and increase to between 3.0--4.0\% when using doubly-deuterated column densities. This represents an enhancement in the D/H ratio by a factor of 2 in IRAS 16293A and more than a factor of 3 in IRAS 16293B. These D/H values are consistent with those found for methanol, ketene, formic acid, formaldehyde, formamide, isocyanic acid, cyanamide and formaldehyde in this object with PILS data ($\sim$2--3\%; J{\o}rgensen et al. subm., \citealt{Coutens2016}, \citealt{Coutens2018}, \citealt{Persson2018}). Those species were also found to have a $^{12}$C/$^{13}$C ratio similar to the ISM, in contrast to species with larger D/H ratios (5--6\%; e.g. ethanol, methyl formate and glycolaldehyde), which have a lower $^{12}$C/$^{13}$C ratio (25-41; J{\o}rgensen et al. subm.). Methyl cyanide fits in with this pattern.

\subsection{Spatial extent}\label{sec:spatial}
\begin{figure*} 
 \begin{center} 

  \includegraphics[width=15cm, angle=0, clip =true, trim = 5cm 11.2cm 5cm 9cm]{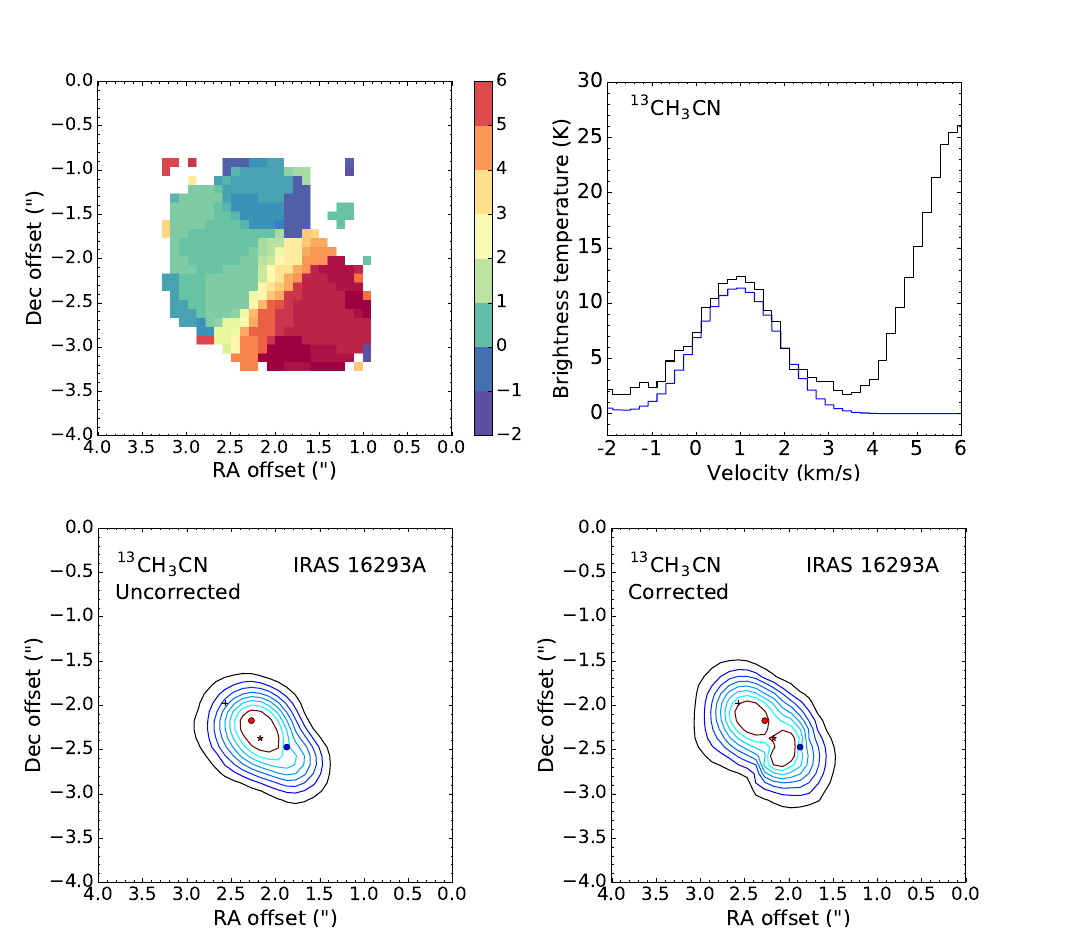}

 \end{center} 
 \caption{{\it Top left:} Velocity map of IRAS 16293A. {\it Top right:} Spectral lines covered over the velocity gradient of IRAS 16293A. {\it Bottom left:} Integrated emission map where the emission for the 338.360\,GHz line of $^{13}$CH$_3$CN is integrated over channels between 0 and 6 km\,s$^{-1}$. {\it Bottom right:} Velocity-corrected integrated emission map of the 338.360\,GHz line of $^{13}$CH$_3$CN where only the line profile at each pixel is integrated over. The red star marks the peak continuum position in the PILS dataset of IRAS 16293A and the black cross marks the offset position where the spectra analysed in this work are extracted from. The red and blue circles mark the positions of the two continuum peaks A1 and A2 respectively, found by previous authors, using coordinates that have been corrected by the rate of position angle change with year determined by \citet{Pech2010}. \label{fig:velmap}} 
 \end{figure*}

To explore the spatial extent of the molecules detected in this work we have produced emission maps for each molecule. A large velocity gradient ($\sim$6\,km\,s$^{-1}$) is observed around IRAS 16293A due to its near edge-on rotating structure \citep{Pineda2012}. This makes producing a channel map by integrating over the entire line profile for a given line difficult as it leads to contamination from other nearby molecular lines due to the range of velocities that have to be integrated over and the crowded spectra observed in hot corinos. In the past, this has limited the spatial information that can be determined for given species around IRAS 16293A. In this paper, however, we use a new method of determining the molecular emission structure around IRAS 16293A. We use a bright transition of methanol, the 7$_{3, 5}$--6$_{4, 4}$ line at 337.519\,GHz, to produce a velocity map of IRAS 16293A (Figure \ref{fig:velmap}, top left). Methanol was chosen because it traces the hot corino emission and has very bright lines which can be used to determine how much the peak velocity of a given line at a given pixel differs from the measured $V_{\rm LSR}$ of the source. We then use this velocity map to determine the centroid channel for a given transition in each pixel, integrating over several channels either side to cover the whole spectral profile. We choose lines that are not blended and integrate over all the channels that contain emission above the RMS of the spectra. We refer to such maps as Velocity-corrected INtegrated Emission (VINE) maps henceforth. Whilst velocity correction techniques have been employed in the past \citep{Matra2014, Marino2016, Yen2016}, this method is fundamentally different, in that it is the first time a velocity correction has been applied to an integrated emission map. 

As an example, the bottom left and right panels of Figure \ref{fig:velmap} show two emission maps of the 339.360\,GHz line of $^{13}$CH$_3$CN. The first map is the emission integrated over 6\,km\,s$^{-1}$. The second emission map has been velocity-corrected, and hence only includes the emission over the line profile of $^{13}$CH$_3$CN. In the first emission map the range of velocities that is integrated over to cover the emission across the A source leads to contamination by a nearby line, shown in the top right panel of Figure \ref{fig:velmap}. Producing a velocity-corrected map means the emission is not contaminated by other lines, and previously unseen structure is resolved as is seen in the second emission map. We recommend using this method for the future analysis of IRAS 16293A molecular emission and other line-rich sources where blending may be an issue.\\

\begin{figure*} 
\begin{center} 

\includegraphics[width=22cm, angle=0, clip =true, trim = 1cm 1.1cm 2cm 2cm]{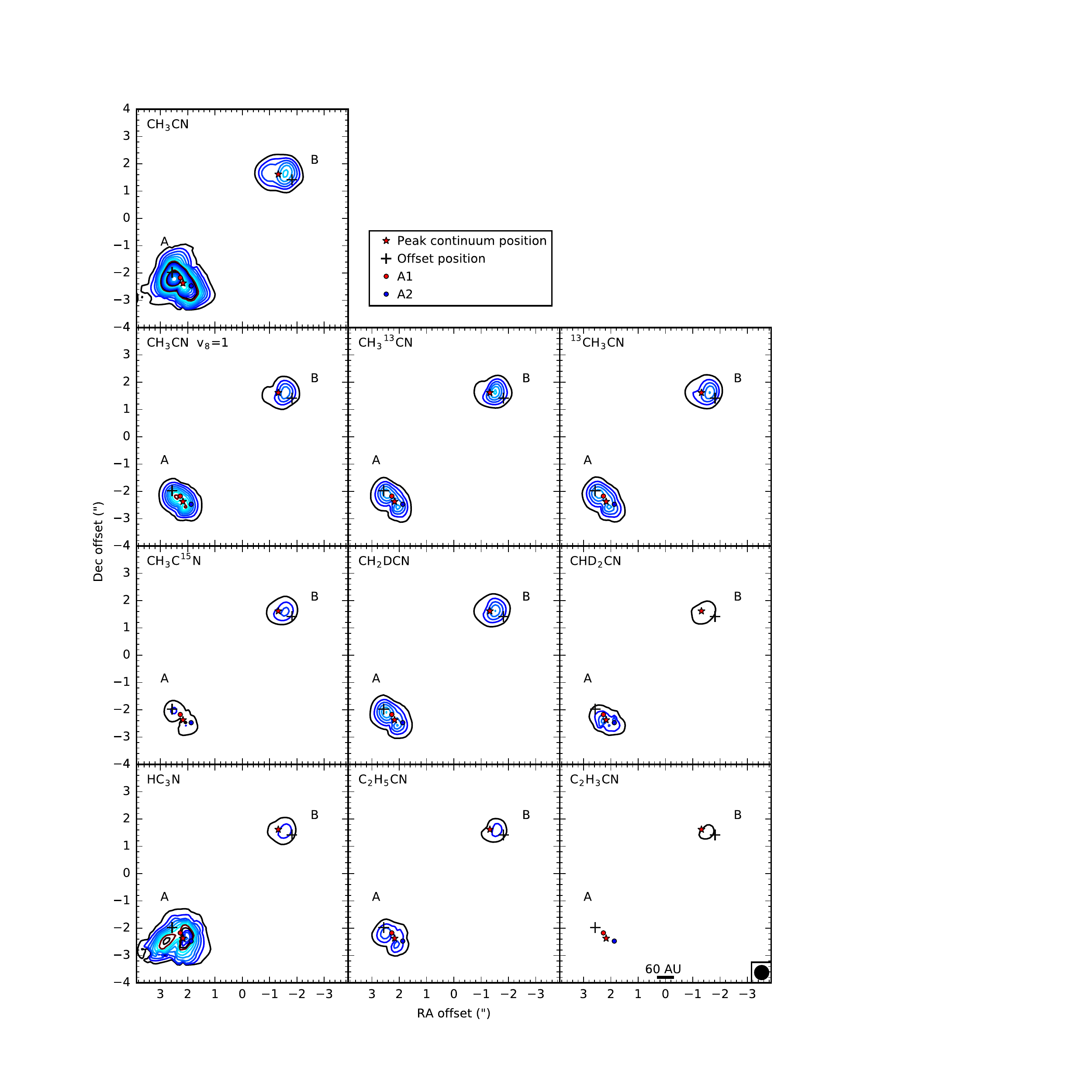}

  \end{center} 
 \caption{Velocity-corrected integrated emission (VINE) maps of the 349.346\,GHz line of CH$_3$CN v=0, the 350.168\,GHz line of CH$_3$CN v$_8$=1, the 339.366\,GHz line of $^{13}$CH$_3$CN, the 349.254\,GHz of CH$_3^{13}$CN, the 338.940\,GHz of CH$_3$C$^{15}$N, the 347.110\,GHz of CH$_2$DCN, the 362.520\,GHz of CHD$_2$CN, the 329.235\,GHz of C$_2$H$_5$CN, the 341.564\,GHz of C$_2$H$_3$CN and the 336.520\,GHz of HC$_3$N. The axes show the position offset from the phase centre of the observations. Contour levels start at 80\,mJy\,km\,s$^{-1}$ and increase in steps of 11\,mJy\,km\,s$^{-1}$. The red stars mark the peak continuum position in the PILS dataset of IRAS 16293A and IRAS 16293B. The black crosses mark the offset positions where the spectra analysed in this work are extracted from. The red and blue circles mark the positions of the two continuum peaks A1 and A2 respectively, found by previous authors, using coordinates that have been corrected by the rate of position angle change with year determined by \citet{Pech2010}.\label{fig:map}} 
 \end{figure*}

Figure \ref{fig:map} shows the VINE map for all of the species detected in this work using unblended bright lines. All of the species trace the hot gas surrounding sources A and B, with higher upper energy lines tracing more compact, hotter gas in both sources. Emission around IRAS 16293A has a larger spatial distribution, being on average 1.5 times larger than source B. The larger distribution of methyl cyanide compared to its isotopologues in both sources can be attributed to the difference in line strength between the emission of these molecules.
 
Emission around IRAS 16293B peaks to the west of the peak continuum position (red star in the figure) due to opacity effects. For most species emission around IRAS 16293A peaks in two locations either side of the continuum peak position (red star in the figure). A zoomed in version of this double peak structure can be found in Figure \ref{fig:nmap_zoom} in Appendix \ref{sec:appen2}. We do not observe a similar double-peaked structure in the PILS continuum results in the left panel of Figure \ref{fig:sourceA} as the emission is optically thick at the observed frequency. Observations at longer wavelengths by \citet{Wootten1989}, however, did detect two continuum sources, A1 and A2, associated with IRAS 16293A. \citet{Chandler2005} and \citet{Pech2010} also detected multiple components associated with IRAS 16293A and observed a change in their coordinates. The red and blue circles on the figure mark the location of A1 and A2 respectively, using coordinates that have been corrected by the rate of position angle change with year determined by \citet{Pech2010}. The small differences observed between the two peaks in the molecular emission found in this work compared to the corrected A1 and A2 continuum peaks can be attributed to pointing errors and errors associated with the coordinate correction.  Whilst this double-peaked emission could be due to multiple sources in IRAS 16293A, it is more likely to be due to the rotating toroidal structure around source A which is near edge-on. Such emission has been observed before in inclined rotating structures (e.g. \citealt{Andrews2009,Matra2018}). Furthermore, it has been seen in 3D radiative transfer modelling of IRAS 16293 by \citet{Jacobsen2018}, where double-peaked emission was found when generating synthetic CO isotopologue emission maps using a rotating toroid model.

\section{Discussion} \label{sec:dis} 
\subsection{Relative abundances}

The results from the abundance determination in each source, as well as the isotopic ratios determined for the $^{12}$C/$^{13}$C, $^{14}$N/$^{15}$N and D/H present a picture of two chemically similar sources. In particular, if the relative abundance is determined with respect to methyl cyanide instead of methanol the differences seen between the molecules are minimal, especially when considering the errors of the observations. The emission maps of both sources suggest a difference in the large-scale emission of nitriles. Previous work by \citet{Bisschop2008} on HNCO and methyl cyanide emission indicated that nitrogen-bearing species are more abundant relative to methanol (derived from the CH$_3$$^{18}$OH column density) towards B compared with A, with only 10--20\% arising from source B. Whilst we do not find such a stark difference in abundance, likely due to unaccounted for optical depth, beam dilution effects in the previous estimate, and a larger beam, we do find that the bulk of nitrile emission peaks towards the A source, showing a more extended spatial distribution which is particularly evident in HC$_3$N and methyl cyanide especially when compared to the spatial extent of CH$_3$$^{18}$OH. Such comparisons between sources are also highly dependent on which species the abundance is calculated relative to, as we show in Section \ref{sec:rel}, where abundances computed relative to methanol show a greater difference between A and B. \\

The significant difference that is seen between the sources is the relative abundance of vinyl cyanide, which is not detected towards IRAS 16293A but is detected towards IRAS 16293B. A factor of 9 difference in abundance between the sources indicates a significant difference in the chemistry or physical conditions for this species. This difference in abundance is also seen in the emission maps presented in Section \ref{sec:spatial}, with no vinyl cyanide emission at any position around IRAS 16293A. Some authors have previously suggested that vinyl cyanide could be used as an evolutionary indicator in hot cores (\citealt{Caselli1993}; \citealt{Fontani2007}), with higher abundances relative to ethyl cyanide indicating a more evolved object. Differences in the ratio of vinyl cyanide to ethyl cyanide can be expected at different stages of evolution as their formation pathways are closely linked. Vinyl cyanide forms early in low-mass star formation on dust grains from the hydrogenation of HC$_3$N, however, it is rapidly hydrogenated further to ethyl cyanide. \citet{Garrod2017} find that the vinyl cyanide abundance only becomes significant later during the warm-up phase, when it is formed as a result of the destruction of ethyl cyanide through protonation and subsequent dissociative electronic recombination, producing vinyl cyanide as one of several possible branches. These models suggest that differences in vinyl cyanide abundance between sources either indicate a difference in evolution or at least a difference in the timescale of the warm-up phase. In IRAS 16293B the ethyl cyanide to vinyl cyanide ratio is 5 while in IRAS 16293A it is >53, which if such a ratio can be used as an evolutionary indicator, would suggest that the B source is more evolved or at least has a longer warm-up timescale, having had more time for vinyl cyanide to be formed in the gas. \\

The outflow distribution in IRAS 16293, however, paints a different picture of the evolution of these two sources. At least two major outflows have been observed from IRAS 16293A, a northwest outflow \citep{Kristensen2013, Girart2014} and an east-west bipolar outflow \citep{Yeh2008}. IRAS 16293B, however, shows no clear signs of outflows, which suggests it is the younger, more embedded source. These conflicting indications of evolution make it difficult to determine the exact evolution of either source. There is, however, another scenario which could lead to both observed differences, if both sources have the same physical age and mass but the A source has higher accretion rate. This would lead to outflows from source A and warm-up timescales so short that vinyl cyanide is not efficiently formed. Such differences would lead to a higher luminosity for the A source compared to the B source, which is supported by recent 3D modelling of the envelope, disks and dust filament by \citet{Jacobsen2018}. 

\subsection{Isotopic ratios}
The isotopic ratios presented in Section \ref{sec:isotopic} are not able to put strict constraints on the formation paths of methyl cyanide. They do, however, tell us something about the enrichment of methyl cyanide isotopologues in IRAS 16293. In particular, the higher D/H ratio when determined from the doubly-deuterated form of methyl cyanide, compared to the singly-deuterated form, suggests an enrichment of deuterium in methyl cyanide. A similar enrichment in the D/H ratio is also found in formaldehyde \citep{Persson2018} and methyl formate \citep{Manigand2018}, implying a mechanism for the enrichment of deuterated species. \citet{Oba2016b, Oba2016a} found H--D substitution could occur on the methyl group of ethanol and dimethyl ether on ice grain surfaces. This could be occurring for methyl cyanide. This process seems more enhanced for IRAS 16293B with D/H ratios derived from the doubly-deuterated form even more enhanced compared to the singly-deuterated form. This could imply a longer period of H--D substitution in IRAS 16293B than in IRAS 16293A.

\subsection{Comparison to other sources}

To put the results of this analysis into a wider context it is useful to compare them to similar studies of complex organic molecules in other objects.  The EMoCA survey of complex organic molecules with ALMA in Sagittarius B2(N2) (Sgr B2(N2) henceforth) provides the data to do such a comparison. Sgr B2(N2) is one of the most line rich sources and therefore one of the most studied sources in the galaxy. It is also the only other source where a detailed study of nitrile species has been performed with high-resolution observations. In the top panel of Figure \ref{fig:rel_abun} the abundance of nitriles with respect to methyl cyanide determined in Sgr B2(N2) by \citet{Belloche2016} is plotted as well as the abundances for IRAS 16293A and IRAS 16293B. Similar abundances are found for CH$_3$C$^{15}$N between the three sources, for the other species large differences are seen. Both forms of $^{13}$C methyl cyanide are $\sim$3 times more abundant in Sgr B2(N2) than in IRAS 16293. It is interesting that there are no abundance differences for different functional groups in any of the three sources. The deuteration level also seems to be less in Sgr B2(N2) with a lower abundance of CH$_2$DCN and no detection of doubly deuterated methyl cyanide. The abundance of ethyl cyanide, vinyl cyanide, and cyanoacetylene are also significantly more abundant in Sgr B2(N2) compared to IRAS 16293.  

\section{Summary and conclusions}  \label{sec:con}

This work presents a detailed study of nitriles in IRAS 16293. Using ALMA observations from the PILS survey and LTE spectral modelling, the excitation temperatures and column densities towards both protostellar sources IRAS 16293A and IRAS 16293B are determined. The main findings are:
\begin{itemize}
\item[$\bullet$] Methyl cyanide is detected in both the v=0 and v$_8$=1 states, as well as 5 of its isotopologues, including the detection of CHD$_2$CN for the first time in the ISM. Ethyl cyanide, vinyl cyanide,  and HC$_3$N are also detected.\\

\item[$\bullet$] All species have excitation temperatures in the range of 100\,--160\,K, based on LTE spectral models of their emission. We have also used LTE spectral models to determine the relative abundance of nitriles in IRAS 16293A and IRAS 16293B, with respect to methanol and methyl cyanide. Most nitriles peak in bulk emission towards IRAS 16293A, however, the differences in abundance between A and B on small scales are not significant, particularly when determined with respect to methyl cyanide. The one exception to this trend is vinyl cyanide which is detected in IRAS 16293B but not in IRAS 16293A. Vinyl cyanide is found to be at least 9 times more abundant towards the B source. The large difference in abundance between the sources for vinyl cyanide is due to physical or chemical differences.\\

\item[$\bullet$] The formation paths of ethyl cyanide and vinyl cyanide are closely linked in the models of \citet{Garrod2017}, with their relative abundance being highly dependent on evolutionary stage. In such a scenario, the differences observed in IRAS 16293 would imply the B source is more evolved, having had time for vinyl cyanide to form in the gas, or warm-up timescales are so short in IRAS 16293A that vinyl cyanide is not efficiently formed. If both sources have the same physical age and mass such differences could also be seen if IRAS 16293A has a higher accretion rate. \\

\item[$\bullet$] Comparison of these results to similar studies in Sgr B2(N2) highlight how despite the difference between low-mass sources and Galactic Centre hot cores, differing functional group does not seem to impact the abundance of $^{13}$C isotopologues. \\

\item[$\bullet$] The isotopic ratios determined in this work are similar for both IRAS 16293A and IRAS 16293B. The $^{12}$C/$^{13}$C ratio determined for methyl cyanide are 80$\pm$19 ($^{13}$CH$_3$CN/CH$_3$CN) and 67$\pm$14 (CH$_3^{13}$CN/CH$_3$CN) in IRAS 16293A and 67$\pm$17 ($^{13}$CH$_3$CN/CH$_3$CN) and 80$\pm$20 (CH$_3^{13}$CN/CH$_3$CN) in IRAS 16293B. These values are within the error range of the ISM value of 68$\pm$15 \citep{Milam2005}. The D/H values (1.2$\pm0.3$--1.5$\pm0.3$\%) for both sources when using singly-deuterated column densities are consistent with those found for methanol, ketene, formic acid, formaldehyde, formamide and isocyanic acid ($\sim$2\%) in this object by previous authors (J{\o}rgensen et al. subm., \citealt{Coutens2016}) which also have a $^{12}$C/$^{13}$C ratio similar to the ISM. This contrasts with ethanol, methyl formate, and glycolaldehyde which exhibit lower  $^{12}$C/$^{13}$C (25--41) ratios and larger D/H ratios (5--6\%). The D/H values derived from doubly-deuterated column densities are a factor of 2 higher in IRAS 16293A and more than a factor of 3 higher in IRAS 16293B. A similar enhancement in the D/H ratio is also found in formaldehyde \citep{Persson2018} and methyl formate \citep{Manigand2018}, suggesting that a mechanism of enrichment occurs, such as H--D substitution.\\

\item[$\bullet$] In this work we have also explored the spatial extent of nitrile organics. We have presented a new technique to explore the spatial structures in sources with high molecular line density and large velocity gradients --- Velocity-corrected INtegrated Emission (VINE) maps. Using this novel technique, we resolve two peaks in the molecular emission around source A for the first time, as has been found previously in continuum observations (see \citealt{Chandler2005, Pech2010, Wootten1989}). This double peak structure is likely a consequence of the rotating toroid structure around source A, and not an indication of multiple sources in source A.
\end{itemize}

Whilst most abundances presented in this work are similar between IRAS 16293A and IRAS 16293B, the abundances of vinyl cyanide indicate possible differences in the evolution of these two sources, the reason for their chemical differentiation is far from clear, and further chemical models are needed that fully reproduce their relative abundances. Further higher angular resolution observations are also needed to understand the double-peaked emission observed in the A source.  The number of nitriles and their isotopologues detected in IRAS 16293 highlight the rich chemistry which is observed in this object. Observing the emission of many different species in the same family of molecules can build up a detailed picture of the chemical similarities and differences between IRAS 16293A and IRAS 16293B, as well as probe the physical structures observed in both sources.\\

\begin{acknowledgements}
The authors would like to acknowledge the European Union whose support has been essential to this research. In particular a European Research Council (ERC) grant, under the Horizon 2020 research and innovation programme (grant agreement No. 646908) through ERC Consolidator Grant "S4F" to J.K.J. The group of EvD acknowledges ERC Advanced Grant "CHEMPLAN" (grant agreement No 291141). AC postdoctoral grant is funded by the ERC Starting Grant 3DICE (grant agreement 336474). MVP postdoctoral position is funded by the ERC consolidator grant 614264. S.F.W. acknowledges financial support from a CSH fellowship. Research at the Centre for Star and Planet Formation is funded by the Danish National Research Foundation. This paper makes use of the following ALMA data: ADS/JAO.ALMA\#2013.1.00278.S. ALMA is a partnership of ESO (representing its member states), NSF (USA) and NINS (Japan), together with NRC (Canada) and NSC and ASIAA (Taiwan), in cooperation with the Republic of Chile. The Joint ALMA Observatory is operated by ESO, AUI/NRAO and NAOJ.

\end{acknowledgements}

%-------------------------------------------------------------------

\bibliography{nitrogen}
\bibliographystyle{aa}

\newpage
\appendix

\section{Spectra}
Spectra of the 9 brightest lines of CH$_3$CN v$_8$=1, $^{13}$CH$_3$CN, CH$_3$$^{13}$CN, CH$_3$C$^{15}$N, CH$_2$DCN, CHD$_2$CN, C$_2$H$_{3}$CN, C$_2$H$_{5}$CN, and HC$_{3}$N.

\begin{figure*} [h]
\begin{center} 
\includegraphics[width=16.5cm, angle=0, clip =true, trim = 4cm 11.6cm 4cm 9.3cm]{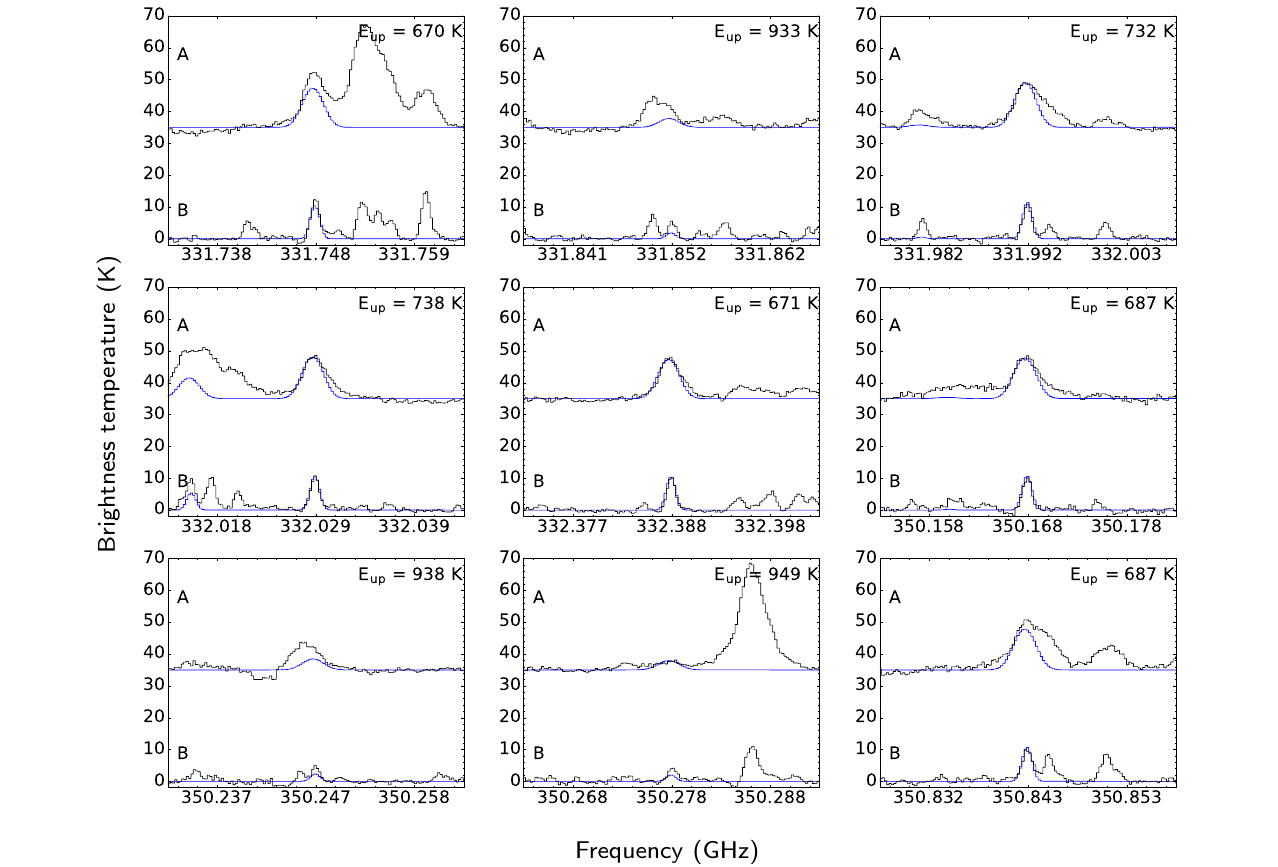}
 \end{center} 
 \vspace{-0.5cm}
 \caption{The 9 brightest lines of CH$_3$CN v$_8$=1 detected in IRAS 16293A and IRAS 16293B overlaid with the LTE spectral model of the emission in each source. Source A is offset by 35 K on the y-axis.\label{fig:ch3cnv8=1}} 
 \end{figure*}  
 
 \begin{figure*}[h] 
 \begin{center}
 \includegraphics[width=16.5cm, angle=0, clip =true, trim = 4cm 11.6cm 4cm 9.3cm]{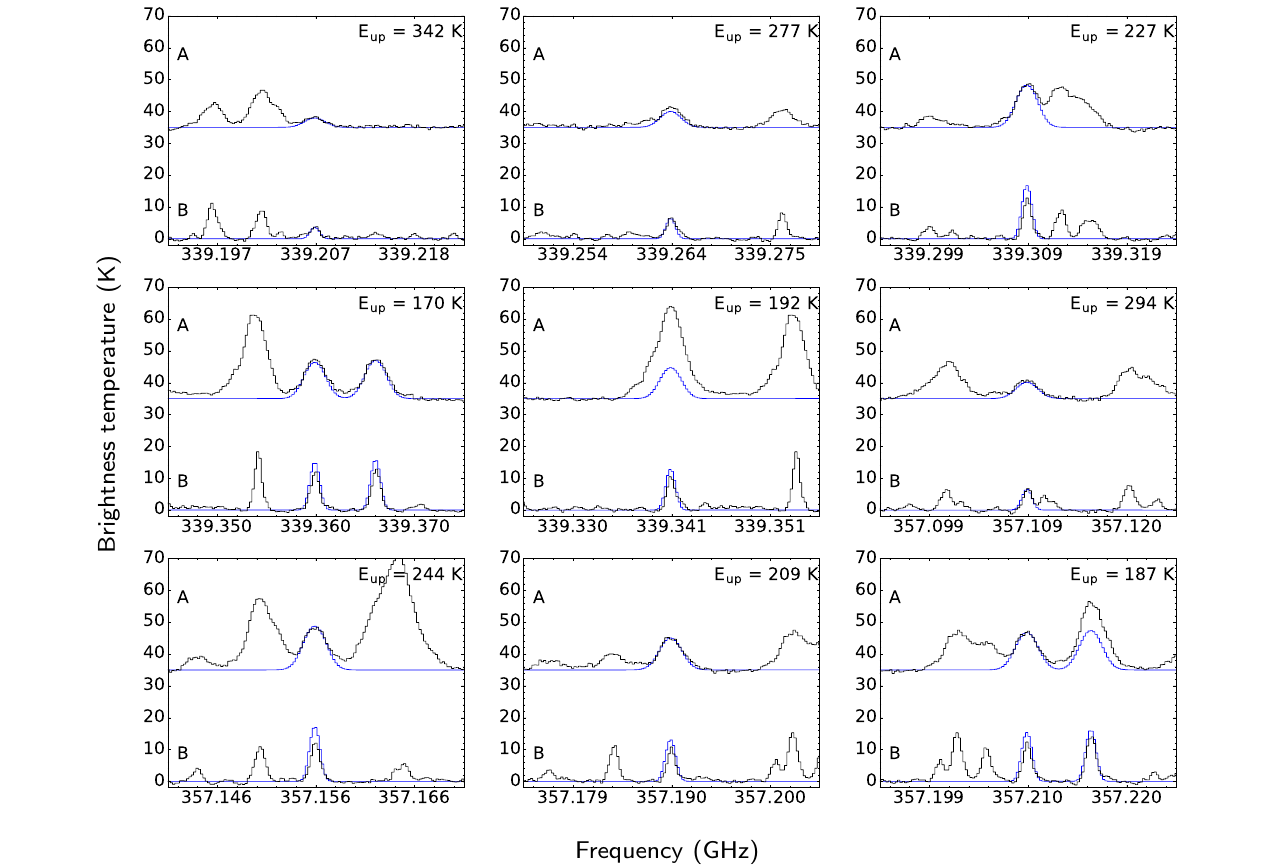}

 \end{center} 
  \vspace{-0.5cm}
 \caption{The 9 brightest lines of $^{13}$CH$_3$CN detected in IRAS 16293A and IRAS 16293B overlaid with an LTE spectral model of the emission in each source. Source A is offset by 35 K on the y-axis. \label{fig:13ch3cn}} 
 \end{figure*}

  \begin{figure*} [h]
 \begin{center} 
  \includegraphics[width=16.5cm, angle=0, clip =true, trim = 4cm 11.6cm 4cm 9.3cm]{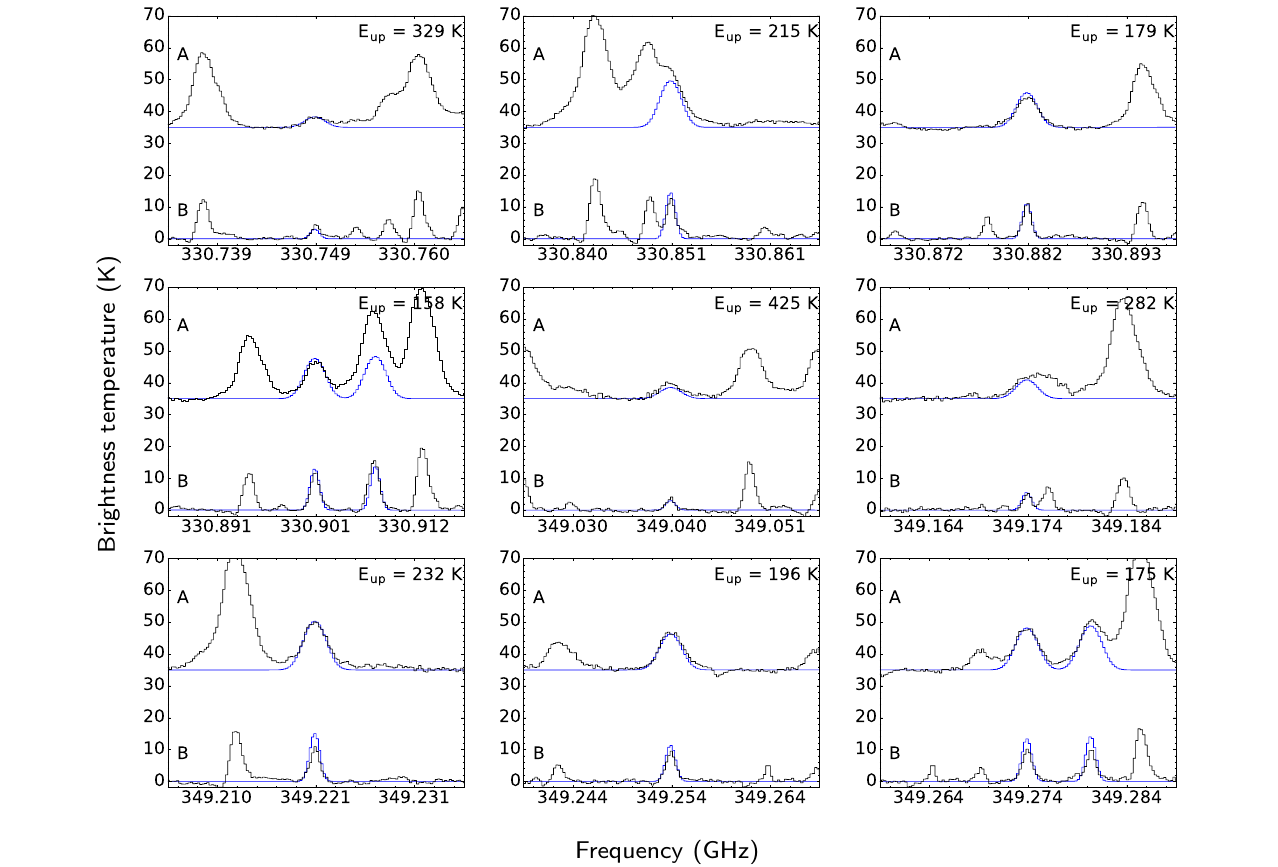}
 \end{center} 
  \vspace{-0.5cm}
 \caption{The 9 brightest lines of CH$_3$$^{13}$CN detected in IRAS 16293A and IRAS 16293B overlaid with an LTE spectral model of the emission in each source. Source A is offset by 35 K on the y-axis. \label{fig:CH313CN}} 
 \end{figure*} 
   
 \begin{figure*} [h]
 \begin{center} 
  \includegraphics[width=16.5cm, angle=0, clip =true, trim = 4cm 11.3cm 4cm 9.3cm]{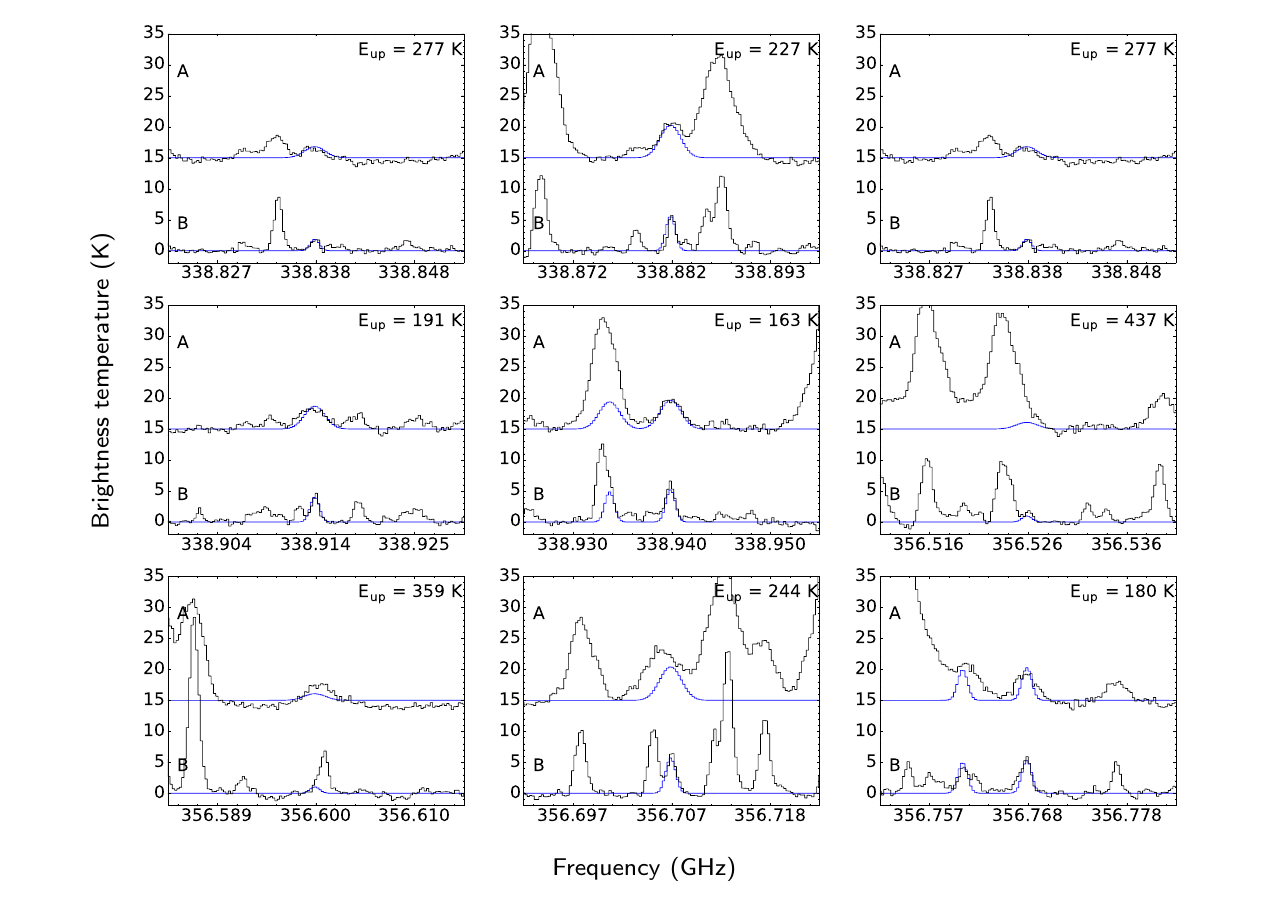}

 \end{center} 
  \vspace{-0.5cm}
 \caption{The 9 brightest lines of CH$_3$C$^{15}$N detected in IRAS 16293A and IRAS 16293B overlaid with an LTE spectral model of the emission in each source. Source A is offset by 35 K on the y-axis. \label{fig:ch3c15n}} 
 \end{figure*}

 \begin{figure*} [h]
 \begin{center} 
  \includegraphics[width=16.5cm, angle=0, clip =true, trim = 4cm 9.8cm 4cm 9.3cm]{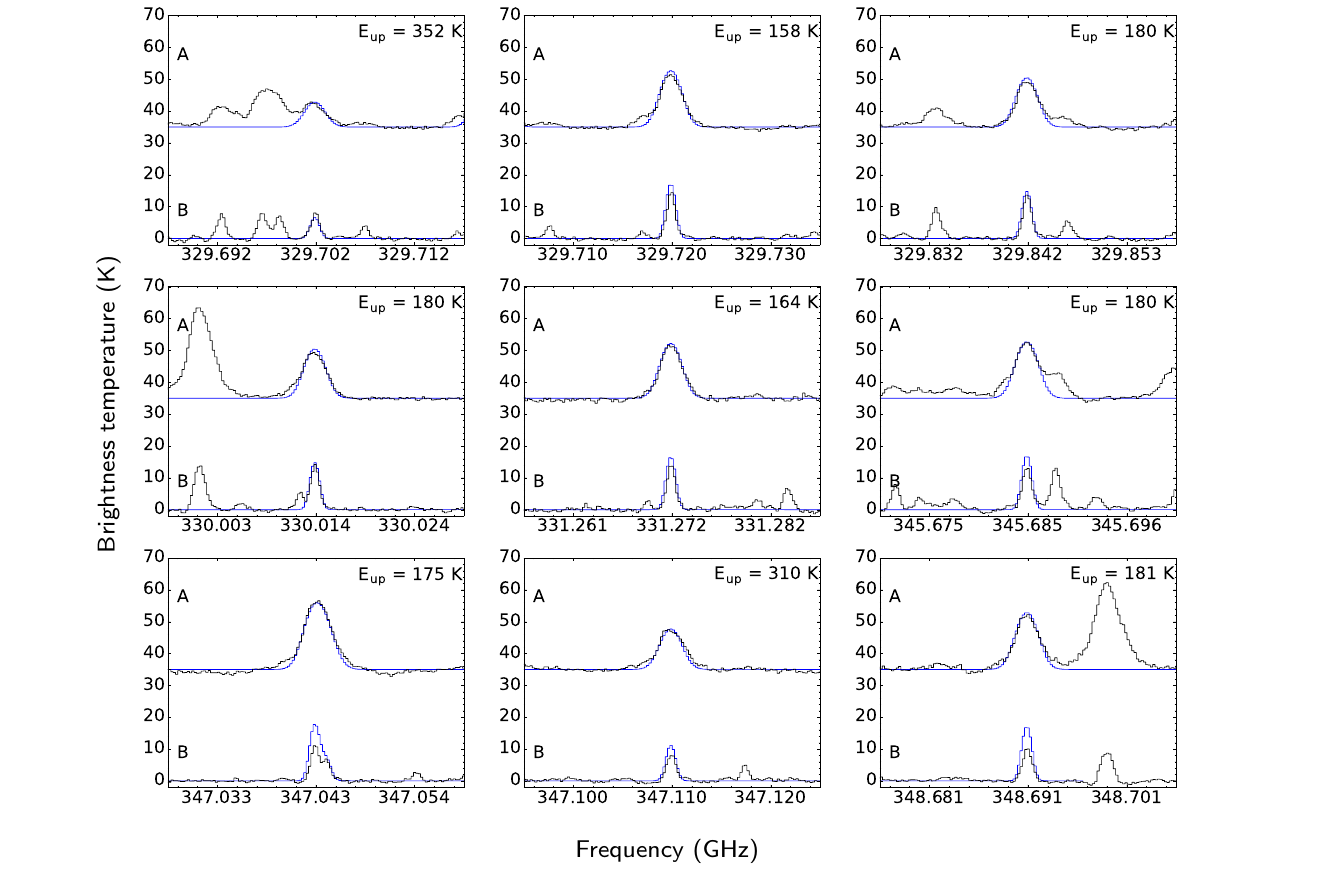}
 \end{center} 
  \vspace{-0.5cm}
 \caption{The 9 brightest lines of CH$_2$DCN detected in IRAS 16293A and IRAS 16293B overlaid with an LTE spectral model of the emission in each source. Source A is offset by 35 K on the y-axis. \label{fig:}} 
 \end{figure*}

 \begin{figure*}[h]
 \begin{center} 
  \includegraphics[width=16.5cm, angle=0, clip =true, trim = 4cm 11.6cm 4cm 9.3cm]{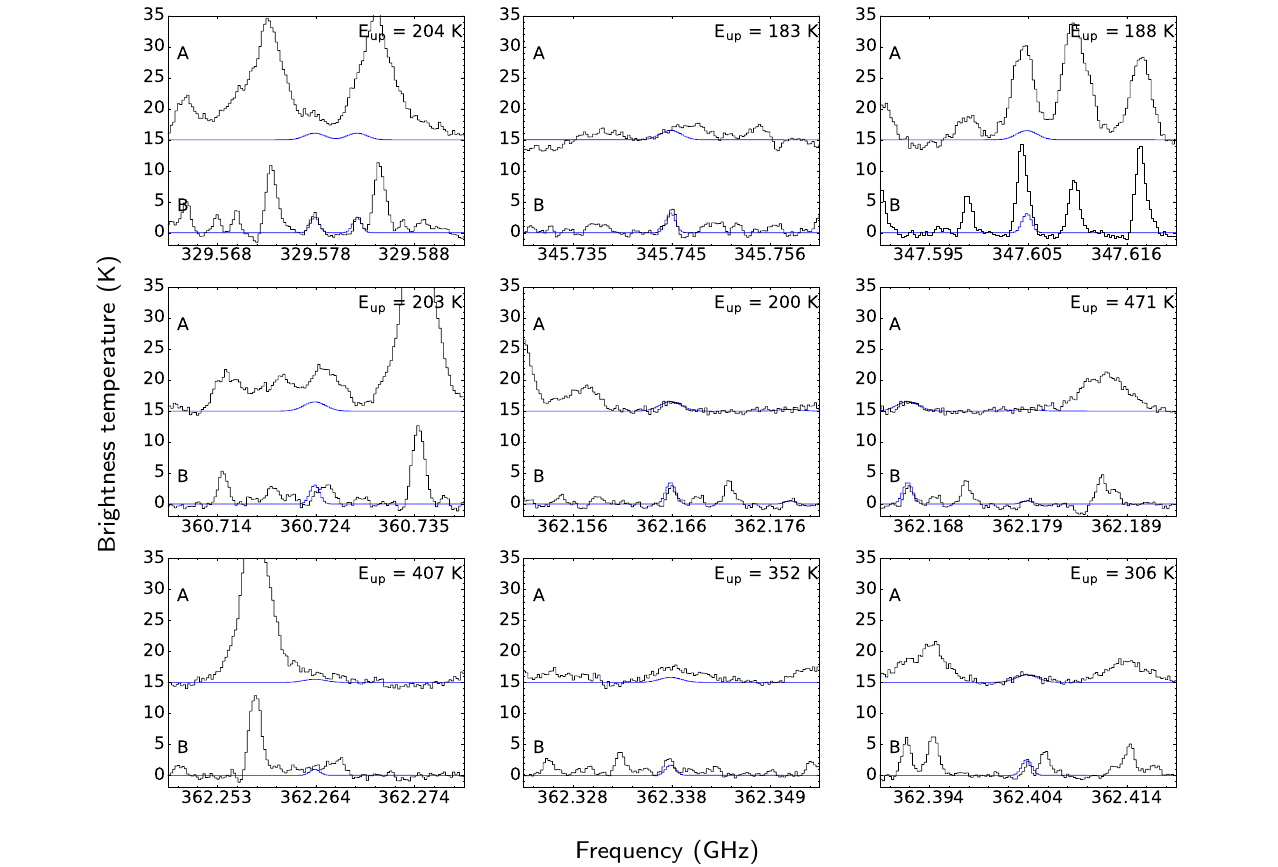}

 \end{center} 
  \vspace{-0.5cm}
 \caption{The 9 brightest lines of CHD$_2$CN detected in IRAS 16293A and IRAS 16293B overlaid with an LTE spectral model of the emission in each source. Source A is offset by 35 K on the y-axis. \label{fig:}} 
 \end{figure*}

 \clearpage
\begin{figure*}[h] 
\begin{center} 
  \includegraphics[width=16.5cm, angle=0, clip =true, trim = 4cm 11.3cm 4cm 9.3cm]{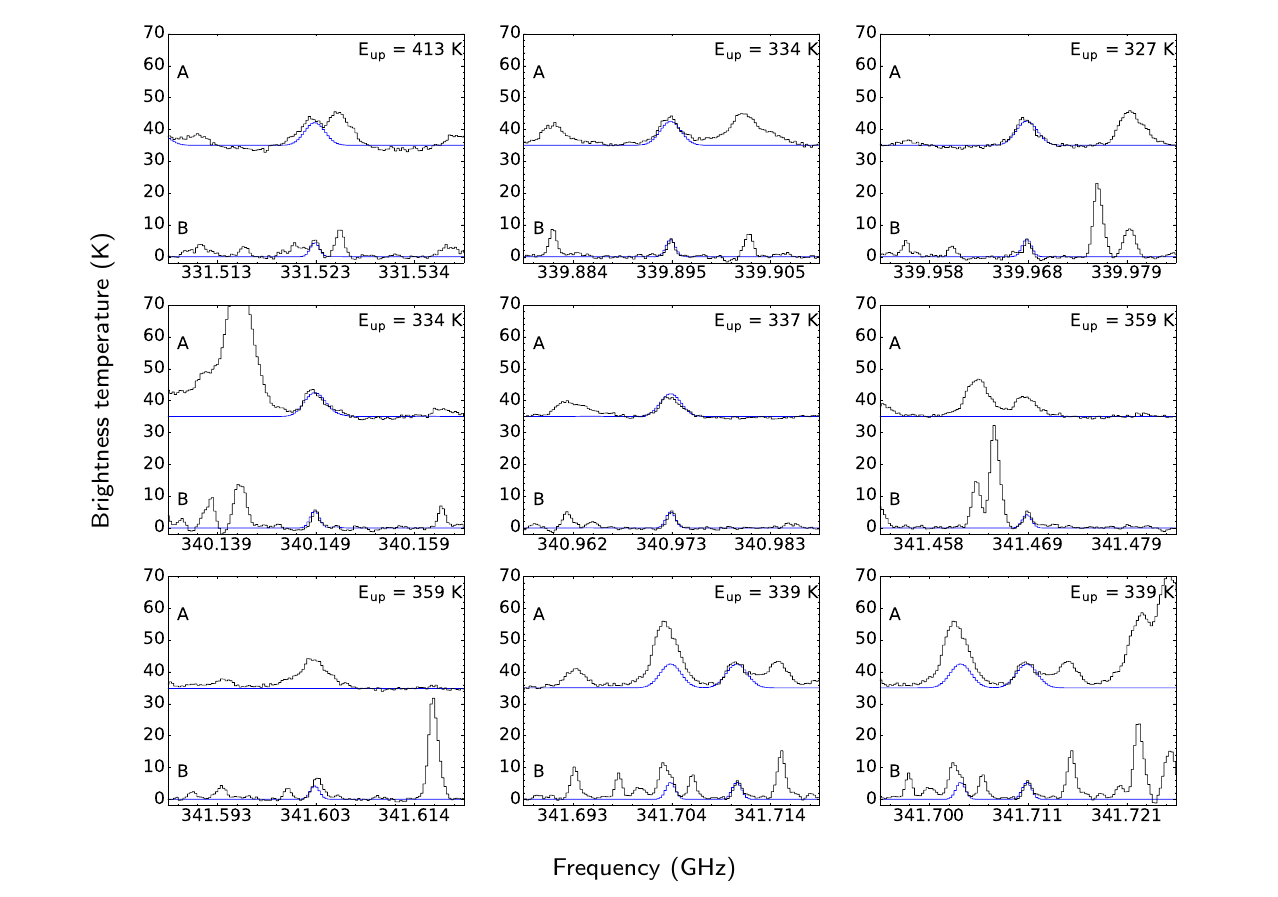}
 \end{center} 
  \vspace{-0.5cm}
 \caption{The 9 brightest lines of C$_2$H$_5$CN detected in IRAS 16293A and IRAS 16293B overlaid with an LTE spectral model of the emission in each source. Source A is offset by 35 K on the y-axis. \label{fig:c2h5cn}} 
 \end{figure*} 

 \begin{figure*}[h] 
 \begin{center} 
  \includegraphics[width=16.5cm, angle=0, clip =true, trim = 4cm 11.6cm 4cm 9.3cm]{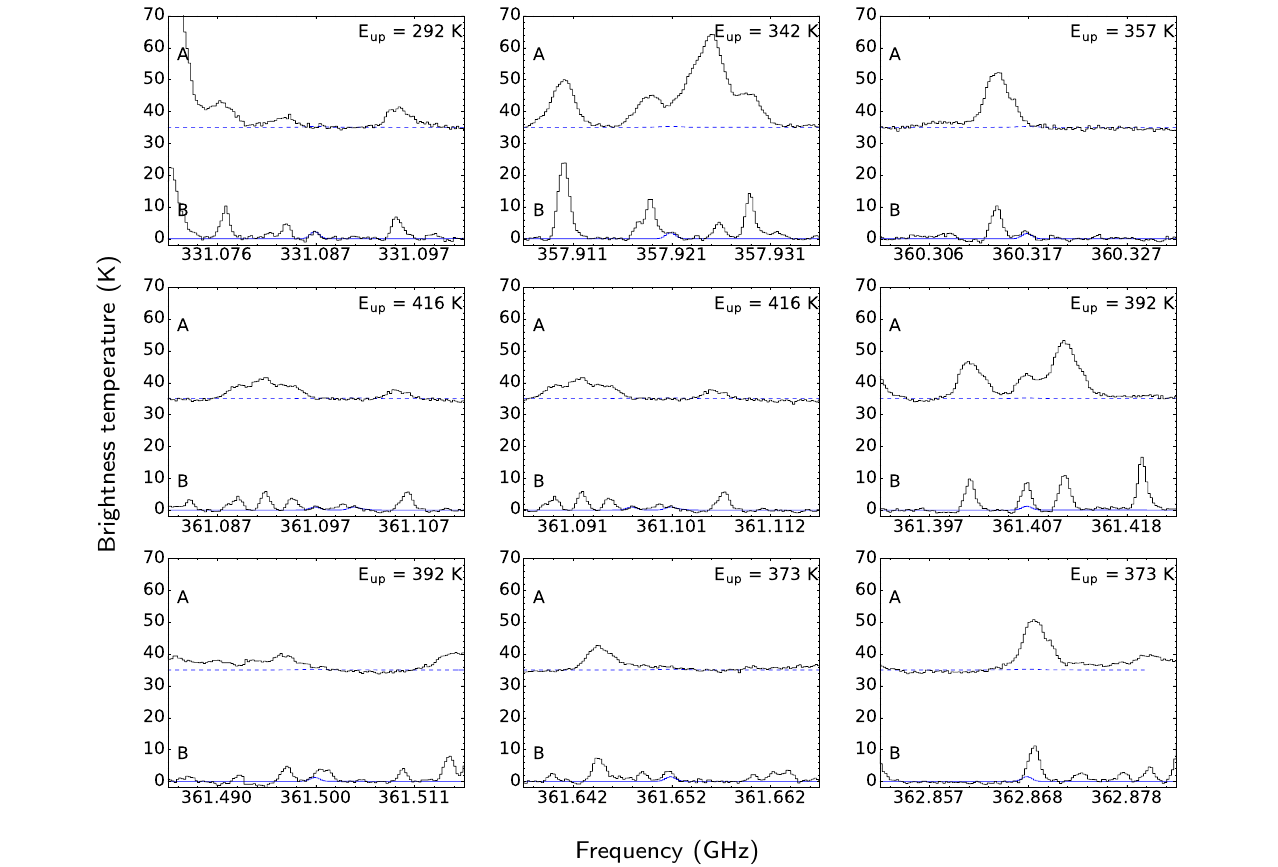}
  \end{center} 
  \vspace{-0.5cm}
 \caption{The 9 brightest lines of C$_2$H$_3$CN detected in IRAS 16293A and IRAS 16293B overlaid with an LTE spectral model of the emission in each source. Source A is offset by 35 K on the y-axis. \label{fig:c2h3cn}} 
 \end{figure*} 

 \begin{figure*}[h] 
 \begin{center} 

  \includegraphics[width=10.5cm, angle=0, clip =true, trim = 3cm 10.7cm 4cm 8cm]{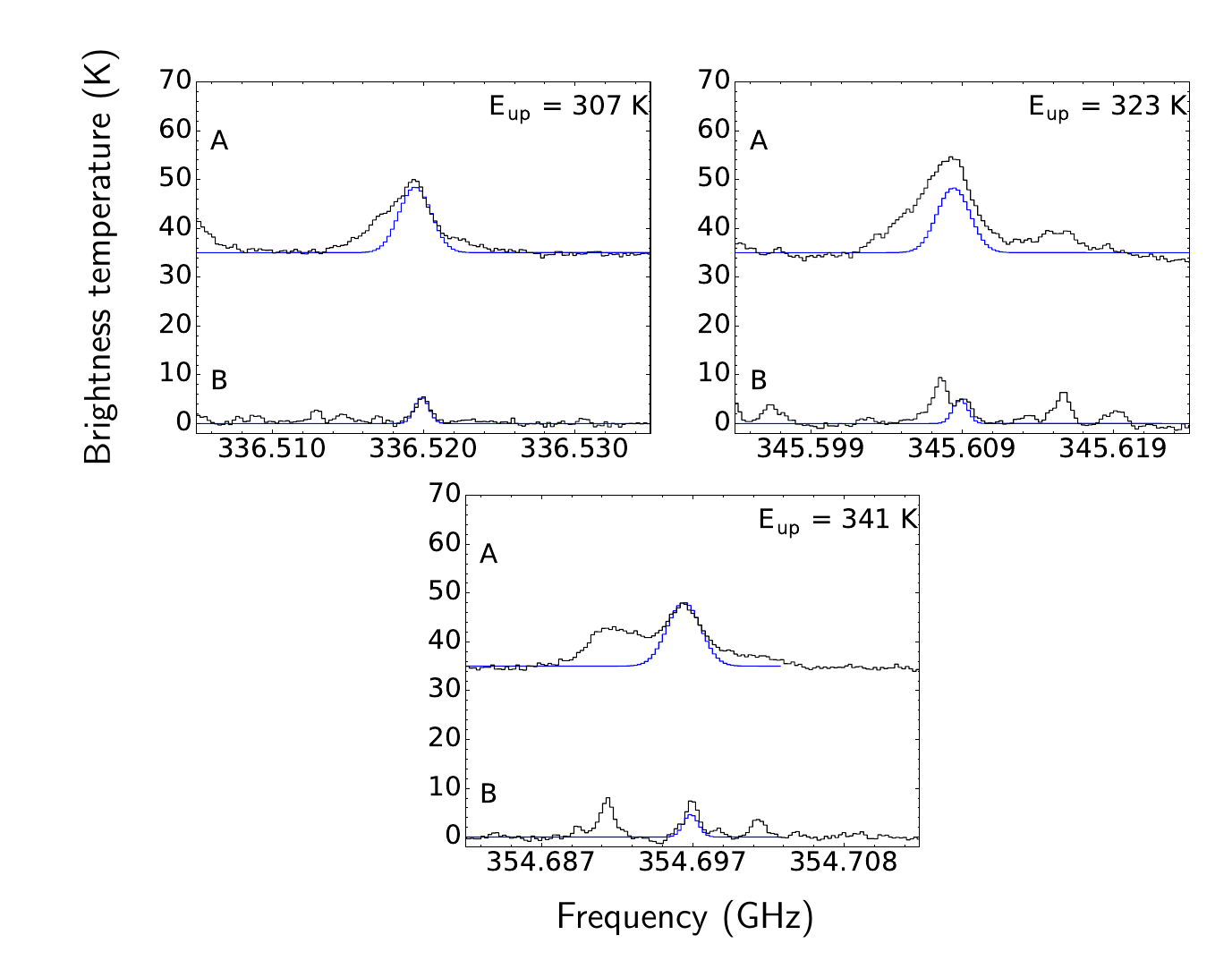}

 \end{center} 
  \vspace{-0.5cm}
 \caption{The 3 lines of HC$_3$N detected in IRAS 16293A and IRAS 16293B overlaid with an LTE spectral model of the emission in each source. Source A is offset by 35 K on the y-axis. \label{fig:hc3n}} 
 \end{figure*}[h] 
\section{Transitional information} \label{sec:trans}
Transitional information for all of the lines detected (>3$\sigma$) in this work.
\begin{table*}[h]
\caption{Detected lines of CH$_3$CN v$_8$=1 \label{tab.lines}}
\begin{tabular}{c c c c c c }
\hline
\hline
Transitions & Frequency & $E_\mathrm{up}$ & $A_\mathrm{ij}$ & Blended A$^a$& Blended B$^a$ \\
            & (MHz) & (K) & (s$^{-1}$) & & \\
\hline						
 18 -1 3 $-$ 17 1 3 & 331748.52& 670& 3.14$\times$10$^{-3}$&Y&Y\\  
 18 -2 2 $-$ 17 2 2 & 331992.35& 732& 3.12$\times$10$^{-3}$&Y&N\\  
 18 2 2 $-$ 17 -2 2 & 331992.35& 732& 3.12$\times$10$^{-3}$&Y&N\\  
 18 0 2 $-$ 17 0 2 & 332015.82& 677& 3.16$\times$10$^{-3}$&Y&Y\\  
 18 -4 3 $-$ 17 4 3 & 332028.76& 738& 3.00$\times$10$^{-3}$&Y&N\\  
 18 4 3 $-$ 17 -4 3 & 332028.76& 738& 3.00$\times$10$^{-3}$&Y&N\\  
 18 2 3 $-$ 17 2 3 & 332107.65& 679& 3.12$\times$10$^{-3}$&Y&Y\\  
 18 1 3 $-$ 17 -1 3 & 332387.75& 671& 3.16$\times$10$^{-3}$&N&N\\  
 19 1 3 $-$ 18 -1 3 & 350168.10& 687& 3.70$\times$10$^{-3}$&N&N\\  
 19 -2 2 $-$ 18 2 2 & 350423.62& 749& 3.68$\times$10$^{-3}$&Y&N\\  
 19 2 2 $-$ 18 -2 2 & 350423.62& 749& 3.68$\times$10$^{-3}$&Y&N\\  
 19 0 2 $-$ 18 0 2 & 350444.91& 693& 3.72$\times$10$^{-3}$&Y&Y\\  
 19 -4 3 $-$ 18 4 3 & 350465.65& 754& 3.56$\times$10$^{-3}$&Y&Y\\  
 19 4 3 $-$ 18 -4 3 & 350465.65& 754& 3.56$\times$10$^{-3}$&Y&Y\\  
 19 -1 3 $-$ 18 1 3 & 350842.67& 687& 3.72$\times$10$^{-3}$&Y&Y\\

\hline
\end{tabular}
\tablefoot{$^a$Y means line is blended with another molecule.}
\end{table*} 
\begin{table*}[h]
\caption{Detected lines of $^{13}$CH$_3$CN \label{tab.lines}}
\begin{tabular}{c c c c c c}
\hline
\hline
Transitions & Frequency & $E_\mathrm{up}$ & $A_\mathrm{ij}$ & Blended A$^a$& Blended B$^a$  \\
            & (MHz) & (K) & (s$^{-1}$) & & \\
\hline	
 19 6 $-$ 18 6 & 339137.34& 420& 3.06$\times$10$^{-3}$&Y&Y\\ 
 19 5 $-$ 18 5 & 339207.24& 342& 3.17$\times$10$^{-3}$&Y&Y\\ 
 19 4 $-$ 18 4 & 339264.47& 277& 3.25$\times$10$^{-3}$&N&N\\  
 19 3 $-$ 18 3 & 339309.01& 227& 3.32$\times$10$^{-3}$&Y&N\\   
 19 2 $-$ 18 2 & 339340.83& 192& 3.37$\times$10$^{-3}$&N&N\\    
 19 1 $-$ 18 1 & 339359.93& 170& 3.40$\times$10$^{-3}$&N&N\\  
 19 0 $-$ 18 0 & 339366.30& 163& 3.41$\times$10$^{-3}$&N&N\\
 20 6 $-$ 19 6 & 356975.47& 437& 3.61$\times$10$^{-3}$&Y&N\\  
 20 5 $-$ 19 5 & 357049.02& 359& 3.73$\times$10$^{-3}$&Y&Y\\  
 20 4 $-$ 19 4 & 357109.23& 294& 3.82$\times$10$^{-3}$&Y&N\\ 
 20 3 $-$ 19 3 & 357156.09& 244& 3.89$\times$10$^{-3}$&Y&N\\ 
 20 2 $-$ 19 2 & 357189.58& 209& 3.94$\times$10$^{-3}$&Y&N\\ 
 20 1 $-$ 19 1 & 357209.67& 187& 3.97$\times$10$^{-3}$&Y&N\\  
 20 0 $-$ 19 0 & 357216.37& 180& 3.98$\times$10$^{-3}$&N&N\\  

\hline
\end{tabular}
\tablefoot{$^a$Y means line is blended with another molecule.}
\end{table*} 
\begin{table*}[h]
\caption{Detected lines of CH$_3$$^{13}$CN \label{tab.lines}}
\begin{tabular}{c c c c c c }
\hline
\hline
 Transitions & Frequency & $E_\mathrm{up}$ & $A_\mathrm{ij}$ & Blended A$^{a}$& Blended B$^{a}$ \\
             & (MHz) & (K) & (s$^{-1}$) & & \\
\hline							
    
 18 6 $-$ 17 6 & 330679.89& 408& 2.80$\times$10$^{-3}$&N&N\\  
 18 5 $-$ 17 5 & 330749.43& 329& 2.91$\times$10$^{-3}$&N&N\\ 
  18 4 $-$ 17 4 & 330806.40& 265& 3.00$\times$10$^{-3}$&Y&Y\\  
 18 3 $-$ 17 3 & 330850.71& 215& 3.07$\times$10$^{-3}$&Y&Y\\ 
 18 2 $-$ 17 2 & 330882.40& 179& 3.12$\times$10$^{-3}$&N&N\\ 
  18 1 $-$ 17 1 & 330901.39& 158& 3.15$\times$10$^{-3}$&N&N\\
 18 0 $-$ 17 0 & 330907.73& 151& 3.16$\times$10$^{-3}$&Y&N\\
 19 6 $-$ 18 6 & 349040.38& 425& 3.34$\times$10$^{-3}$&Y&N\\ 
 19 5 $-$ 18 5 & 349113.77& 346& 3.45$\times$10$^{-3}$&Y&Y\\      
 19 4 $-$ 18 4 & 349173.86& 282& 3.55$\times$10$^{-3}$&Y&Y\\  
 19 3 $-$ 18 3 & 349220.62& 232& 3.62$\times$10$^{-3}$&N&N\\   
 19 2 $-$ 18 2 & 349254.03& 196& 3.67$\times$10$^{-3}$&N&N\\  
 19 1 $-$ 18 1 & 349274.09& 175& 3.71$\times$10$^{-3}$&N&N\\ 
 19 0 $-$ 18 0 & 349280.77& 168& 3.72$\times$10$^{-3}$&Y&N\\  

\hline
\end{tabular}
\tablefoot{$^a$Y means line is blended with another molecule.}
\end{table*} 

\begin{table*}[h]
\caption{Detected lines of CH$_3$C$^{15}$N \label{tab.lines}}
\begin{tabular}{c c c c c c }
\hline
\hline
Transitions & Frequency & $E_\mathrm{up}$ & $A_\mathrm{ij}$ & Blended A$^a$& Blended B$^a$ \\
            & (MHz) & (K) & (s$^{-1}$) & & \\
\hline	
 19 6 $-$ 18 6 & 338710.02& 420& 3.06$\times$10$^{-3}$&Y&Y\\  
 19 5 $-$ 18 5 & 338780.22& 342& 3.16$\times$10$^{-3}$&Y&N\\  
 19 4 $-$ 18 4 & 338837.68& 277& 3.25$\times$10$^{-3}$&N&N\\  
 19 3 $-$ 18 3 & 338882.39& 227& 3.32$\times$10$^{-3}$&Y&Y\\  
 19 2 $-$ 18 2 & 338914.34& 191& 3.36$\times$10$^{-3}$&Y&Y\\  
 19 1 $-$ 18 1 & 338933.54& 170& 3.39$\times$10$^{-3}$&Y&Y\\  
 19 0 $-$ 18 0 & 338939.93& 163& 3.40$\times$10$^{-3}$&N&N\\  
 20 6 $-$ 19 6 & 356525.88& 437& 3.61$\times$10$^{-3}$&Y&Y\\  
 20 5 $-$ 19 5 & 356599.73& 359& 3.72$\times$10$^{-3}$&Y&Y\\  
 20 4 $-$ 19 4 & 356660.20& 294& 3.81$\times$10$^{-3}$&Y&Y\\  
 20 3 $-$ 19 3 & 356707.25& 244& 3.88$\times$10$^{-3}$&Y&Y\\  
 20 2 $-$ 19 2 & 356740.87& 208& 3.93$\times$10$^{-3}$&Y&Y\\  
 20 1 $-$ 19 1 & 356761.05& 187& 3.96$\times$10$^{-3}$&Y&Y\\  
 20 0 $-$ 19 0 & 356767.77& 180& 3.97$\times$10$^{-3}$&N&N\\  
 
\hline
\end{tabular}
\tablefoot{$^a$Y means line is blended with another molecule.}
\end{table*} 
\begin{table*}[h]
\caption{Detected lines of CH$_2$DCN \label{tab.lines}}
\begin{tabular}{c c c c c c }
\hline
\hline
Transitions & Frequency & $E_\mathrm{up}$ & $A_\mathrm{ij}$ & Blended A$^a$& Blended B$^a$ \\
            & (MHz) & (K) & (s$^{-1}$) & & \\
\hline	
CH2DCN 19 8 11 $-$ 18 8 10 & 329548.11& 503& 2.57$\times$10$^{-3}$&N&N\\  
CH2DCN 19 8 12 $-$ 18 8 11 & 329548.11& 503& 2.57$\times$10$^{-3}$&N&N\\  
CH2DCN 19 7 13 $-$ 18 7 12 & 329630.28& 422& 2.70$\times$10$^{-3}$&Y&N\\  
CH2DCN 19 7 12 $-$ 18 7 11 & 329630.28& 422& 2.70$\times$10$^{-3}$&Y&N\\  
CH2DCN 19 6 13 $-$ 18 6 12 & 329702.08& 352& 2.81$\times$10$^{-3}$&Y&N\\  
CH2DCN 19 6 14 $-$ 18 6 13 & 329702.08& 352& 2.81$\times$10$^{-3}$&Y&N\\  
CH2DCN 19 0 19 $-$ 18 0 18 & 329720.02& 158& 3.12$\times$10$^{-3}$&N&N\\  
CH2DCN 19 5 14 $-$ 18 5 13 & 329763.98& 293& 2.91$\times$10$^{-3}$&N&N\\  
CH2DCN 19 5 15 $-$ 18 5 14 & 329763.98& 293& 2.91$\times$10$^{-3}$&N&N\\  
CH2DCN 19 4 16 $-$ 18 4 15 & 329817.09& 245& 2.99$\times$10$^{-3}$&Y&Y\\  
CH2DCN 19 4 15 $-$ 18 4 14 & 329817.10& 245& 2.99$\times$10$^{-3}$&Y&Y\\  
CH2DCN 19 2 18 $-$ 18 2 17 & 329842.21& 180& 3.09$\times$10$^{-3}$&N&N\\  
CH2DCN 19 3 17 $-$ 18 3 16 & 329864.19& 207& 3.05$\times$10$^{-3}$&Y&N\\  
CH2DCN 19 3 16 $-$ 18 3 15 & 329866.11& 207& 3.05$\times$10$^{-3}$&Y&N\\  
CH2DCN 19 2 17 $-$ 18 2 16 & 330013.79& 180& 3.10$\times$10$^{-3}$&Y&Y\\  
CH2DCN 19 1 18 $-$ 18 1 17 & 331271.79& 164& 3.16$\times$10$^{-3}$&N&N\\  
CH2DCN 20 1 20 $-$ 19 1 19 & 345685.38& 180& 3.60$\times$10$^{-3}$&Y&N\\  
CH2DCN 20 8 13 $-$ 19 8 12 & 346882.40& 520& 3.06$\times$10$^{-3}$&Y&N\\  
CH2DCN 20 8 12 $-$ 19 8 11 & 346882.40& 520& 3.06$\times$10$^{-3}$&Y&N\\  
CH2DCN 20 7 13 $-$ 19 7 12 & 346968.95& 439& 3.20$\times$10$^{-3}$&Y&N\\  
CH2DCN 20 7 14 $-$ 19 7 13 & 346968.95& 439& 3.20$\times$10$^{-3}$&Y&N\\  
CH2DCN 20 0 20 $-$ 19 0 19 & 347043.46& 175& 3.65$\times$10$^{-3}$&Y&Y\\  
CH2DCN 20 6 14 $-$ 19 6 13 & 347044.65& 369& 3.32$\times$10$^{-3}$&Y&Y\\  
CH2DCN 20 6 15 $-$ 19 6 14 & 347044.65& 369& 3.32$\times$10$^{-3}$&Y&Y\\  
CH2DCN 20 5 15 $-$ 19 5 14 & 347110.03& 310& 3.42$\times$10$^{-3}$&N&N\\  
CH2DCN 20 5 16 $-$ 19 5 15 & 347110.03& 310& 3.42$\times$10$^{-3}$&N&N\\  
CH2DCN 20 4 17 $-$ 19 4 16 & 347166.41& 261& 3.50$\times$10$^{-3}$&N&N\\  
CH2DCN 20 4 16 $-$ 19 4 15 & 347166.42& 261& 3.50$\times$10$^{-3}$&N&N\\  
CH2DCN 20 2 19 $-$ 19 2 18 & 347188.25& 197& 3.61$\times$10$^{-3}$&N&N\\  
CH2DCN 20 3 18 $-$ 19 3 17 & 347216.88& 224& 3.57$\times$10$^{-3}$&Y&N\\  
CH2DCN 20 3 17 $-$ 19 3 16 & 347219.36& 224& 3.57$\times$10$^{-3}$&Y&N\\  
CH2DCN 20 2 18 $-$ 19 2 17 & 347388.25& 197& 3.62$\times$10$^{-3}$&N&N\\  
CH2DCN 20 1 19 $-$ 19 1 18 & 348690.94& 181& 3.69$\times$10$^{-3}$&N&N\\

\hline
\end{tabular}
\tablefoot{$^a$Y means line is blended with another molecule.}
\end{table*} 
\begin{table*}
\caption{Detected lines of CHD$_2$CN \label{tab.lines}}
\begin{tabular}{c c c c c c }
\hline
\hline
 Transitions & Frequency & $E_\mathrm{up}$ & $A_\mathrm{ij}$ & Blended A$^a$& Blended B$^a$ \\
            & (MHz) & (K) & (s$^{-1}$) & & \\
\hline	
CHD2CN 20 0 20 $-$ 19 0 19 & 329318.90& 166& 3.12$\times$10$^{-3}$&N&N\\  
CHD2CN 20 5 16 $-$ 19 5 15 & 329476.66& 272& 2.93$\times$10$^{-3}$&Y&Y\\  
CHD2CN 20 5 15 $-$ 19 5 14 & 329476.66& 272& 2.93$\times$10$^{-3}$&Y&Y\\  
CHD2CN 20 2 19 $-$ 19 2 18 & 329526.16& 183& 3.09$\times$10$^{-3}$&Y&Y\\  
CHD2CN 20 4 17 $-$ 19 4 16 & 329528.91& 234& 3.00$\times$10$^{-3}$&Y&Y\\  
CHD2CN 20 4 16 $-$ 19 4 15 & 329528.94& 234& 3.00$\times$10$^{-3}$&Y&Y\\  
CHD2CN 20 3 18 $-$ 19 3 17 & 329578.05& 204& 3.05$\times$10$^{-3}$&Y&N\\  
CHD2CN 20 3 17 $-$ 19 3 16 & 329582.49& 204& 3.05$\times$10$^{-3}$&Y&N\\  
CHD2CN 20 2 18 $-$ 19 2 17 & 329797.26& 183& 3.10$\times$10$^{-3}$&Y&N\\  
CHD2CN 20 1 19 $-$ 19 1 18 & 331070.43& 171& 3.16$\times$10$^{-3}$&Y&Y\\  
CHD2CN 21 1 21 $-$ 20 1 20 & 344345.88& 186& 3.56$\times$10$^{-3}$&Y&N\\  
CHD2CN 21 0 21 $-$ 20 0 20 & 345745.31& 183& 3.61$\times$10$^{-3}$&Y&N\\  
CHD2CN 21 5 16 $-$ 20 5 15 & 345941.09& 289& 3.41$\times$10$^{-3}$&Y&Y\\  
CHD2CN 21 5 17 $-$ 20 5 16 & 345941.09& 289& 3.41$\times$10$^{-3}$&Y&Y\\  
CHD2CN 21 2 20 $-$ 20 2 19 & 345986.91& 200& 3.59$\times$10$^{-3}$&Y&Y\\  
CHD2CN 21 4 18 $-$ 20 4 17 & 345996.67& 250& 3.49$\times$10$^{-3}$&Y&Y\\  
CHD2CN 21 4 17 $-$ 20 4 16 & 345996.71& 250& 3.49$\times$10$^{-3}$&Y&Y\\  
CHD2CN 21 3 19 $-$ 20 3 18 & 346049.48& 221& 3.55$\times$10$^{-3}$&N&N\\  
CHD2CN 21 3 18 $-$ 20 3 17 & 346055.14& 221& 3.55$\times$10$^{-3}$&N&N\\  
CHD2CN 21 2 19 $-$ 20 2 18 & 346300.34& 200& 3.60$\times$10$^{-3}$&N&N\\  
CHD2CN 21 1 20 $-$ 20 1 19 & 347605.22& 188& 3.66$\times$10$^{-3}$&Y&Y\\  
CHD2CN 22 1 22 $-$ 21 1 21 & 360724.24& 203& 4.10$\times$10$^{-3}$&Y&Y\\  
CHD2CN 22 0 22 $-$ 21 0 21 & 362166.08& 200& 4.15$\times$10$^{-3}$&N&N\\  
CHD2CN 22 6 17 $-$ 21 6 16 & 362338.39& 352& 3.85$\times$10$^{-3}$&Y&Y\\  
CHD2CN 22 6 16 $-$ 21 6 15 & 362338.39& 352& 3.85$\times$10$^{-3}$&Y&Y\\  
CHD2CN 22 5 18 $-$ 21 5 17 & 362404.16& 306& 3.95$\times$10$^{-3}$&N&N\\  
CHD2CN 22 5 17 $-$ 21 5 16 & 362404.16& 306& 3.95$\times$10$^{-3}$&N&N\\  
CHD2CN 22 2 21 $-$ 21 2 20 & 362445.39& 217& 4.13$\times$10$^{-3}$&N&N\\  
CHD2CN 22 4 19 $-$ 21 4 18 & 362463.16& 268& 4.03$\times$10$^{-3}$&Y&N\\  
CHD2CN 22 4 18 $-$ 21 4 17 & 362463.22& 268& 4.03$\times$10$^{-3}$&Y&N\\  
CHD2CN 22 3 20 $-$ 21 3 19 & 362519.76& 238& 4.09$\times$10$^{-3}$&Y&Y\\  
CHD2CN 22 3 19 $-$ 21 3 18 & 362526.91& 238& 4.09$\times$10$^{-3}$&Y&Y\\  
CHD2CN 22 2 20 $-$ 21 2 19 & 362805.20& 217& 4.14$\times$10$^{-3}$&Y&N\\

\hline
\end{tabular}
\tablefoot{$^a$Y means line is blended with another molecule.}
\end{table*} 
\onecolumn
\begin{longtable}{cccccc}
\caption{Detected lines of C$_2$H$_5$CN \label{tab.lines}}\\
\hline 
\hline 
Transitions & Frequency &$E_\mathrm{up}$&$A_\mathrm{ij}$&Blended A$^{a}$& Blended B$^{a}$\\ 
            & (MHz) & (K) & (s$^{-1}$) & & \\
\hline 
\endfirsthead
{{\bfseries \tablename\ \thetable{} -- continued from previous page}} \\
\hline \hline Transitions &Frequency &$E_\mathrm{up}$&$A_\mathrm{ij}$&Blended A$^{a}$& Blended B$^{a}$ \\ 
            & (MHz) & (K) & (s$^{-1}$) & & \\
\hline 
\endhead
\hline \multicolumn{6}{r}{{Continued on next page}} \\ \hline
\multicolumn{6}{l}{{$^a$Y means line is blended with another molecule.}}

\endfoot
\endlastfoot

 36 3 33 $-$ 35 3 32 & 329234.72& 301& 3.02$\times$10$^{-3}$&N&N\\  
 38 2 37 $-$ 37 2 36 & 331439.54& 318& 3.09$\times$10$^{-3}$&Y&N\\  
 37 12 25 $-$ 36 12 24 & 331484.52& 462& 2.77$\times$10$^{-3}$&Y&Y\\  
 37 12 26 $-$ 36 12 25 & 331484.52& 462& 2.77$\times$10$^{-3}$&Y&N\\  
 37 11 26 $-$ 36 11 25 & 331487.29& 437& 2.83$\times$10$^{-3}$&Y&Y\\  
 37 11 27 $-$ 36 11 26 & 331487.29& 437& 2.83$\times$10$^{-3}$&Y&Y\\  
 37 13 24 $-$ 36 13 23 & 331506.93& 490& 2.72$\times$10$^{-3}$&Y&N\\  
 37 13 25 $-$ 36 13 24 & 331506.93& 490& 2.72$\times$10$^{-3}$&Y&N\\  
 37 10 28 $-$ 36 10 27 & 331523.32& 413& 2.88$\times$10$^{-3}$&Y&Y\\  
 37 10 27 $-$ 36 10 26 & 331523.32& 413& 2.88$\times$10$^{-3}$&Y&Y\\  
 37 14 24 $-$ 36 14 23 & 331549.42& 520& 2.66$\times$10$^{-3}$&N&N\\  
 37 14 23 $-$ 36 14 22 & 331549.42& 520& 2.66$\times$10$^{-3}$&N&N\\  
 37 9 29 $-$ 36 9 28 & 331605.59& 392& 2.92$\times$10$^{-3}$&Y&N\\  
 37 9 28 $-$ 36 9 27 & 331605.59& 392& 2.92$\times$10$^{-3}$&Y&N\\  
 37 2 35 $-$ 36 2 34 & 331662.29& 311& 3.08$\times$10$^{-3}$&N&N\\  
 38 1 37 $-$ 37 1 36 & 331748.53& 318& 3.09$\times$10$^{-3}$&Y&Y\\  
 37 8 29 $-$ 36 8 28 & 331756.44& 373& 2.96$\times$10$^{-3}$&Y&Y\\  
 37 8 30 $-$ 36 8 29 & 331756.44& 373& 2.96$\times$10$^{-3}$&Y&Y\\  
 37 7 31 $-$ 36 7 30 & 332014.73& 357& 3.00$\times$10$^{-3}$&Y&Y\\  
 37 7 30 $-$ 36 7 29 & 332020.59& 357& 3.00$\times$10$^{-3}$&Y&Y\\  
 37 4 34 $-$ 36 4 33 & 332126.30& 321& 3.08$\times$10$^{-3}$&Y&N\\  
 37 6 32 $-$ 36 6 31 & 332428.04& 343& 3.04$\times$10$^{-3}$&Y&N\\  
 37 6 31 $-$ 36 6 30 & 332529.86& 343& 3.05$\times$10$^{-3}$&Y&Y\\  
 37 5 33 $-$ 36 5 32 & 332830.71& 331& 3.08$\times$10$^{-3}$&Y&N\\  
 39 1 39 $-$ 38 1 38 & 333265.90& 323& 3.15$\times$10$^{-3}$&Y&N\\  
 39 0 39 $-$ 38 0 38 & 333274.67& 323& 3.15$\times$10$^{-3}$&Y&N\\  
 37 5 32 $-$ 36 5 31 & 333921.55& 331& 3.11$\times$10$^{-3}$&N&N\\  
 38 3 36 $-$ 37 3 35 & 337347.58& 328& 3.24$\times$10$^{-3}$&Y&Y\\  
 37 4 33 $-$ 36 4 32 & 337445.86& 322& 3.23$\times$10$^{-3}$&Y&Y\\  
 37 3 34 $-$ 36 3 33 & 338142.85& 317& 3.27$\times$10$^{-3}$&Y&Y\\  
 39 2 38 $-$ 38 2 37 & 339894.69& 334& 3.33$\times$10$^{-3}$&N&N\\  
 38 2 36 $-$ 37 2 35 & 339968.22& 327& 3.32$\times$10$^{-3}$&N&N\\  
 39 1 38 $-$ 38 1 37 & 340149.11& 334& 3.34$\times$10$^{-3}$&N&N\\  
 38 12 26 $-$ 37 12 25 & 340432.64& 478& 3.02$\times$10$^{-3}$&N&N\\  
 38 12 27 $-$ 37 12 26 & 340432.64& 478& 3.02$\times$10$^{-3}$&N&N\\  
 38 11 27 $-$ 37 11 26 & 340440.07& 453& 3.08$\times$10$^{-3}$&Y&Y\\  
 38 11 28 $-$ 37 11 27 & 340440.07& 453& 3.08$\times$10$^{-3}$&Y&Y\\  
 38 13 25 $-$ 37 13 24 & 340452.19& 506& 2.97$\times$10$^{-3}$&Y&Y\\  
 38 13 26 $-$ 37 13 25 & 340452.19& 506& 2.97$\times$10$^{-3}$&Y&Y\\  
 38 10 29 $-$ 37 10 28 & 340483.19& 430& 3.13$\times$10$^{-3}$&Y&Y\\  
 38 10 28 $-$ 37 10 27 & 340483.19& 430& 3.13$\times$10$^{-3}$&Y&Y\\  
 38 14 25 $-$ 37 14 24 & 340492.97& 536& 2.91$\times$10$^{-3}$&Y&Y\\  
 38 14 24 $-$ 37 14 23 & 340492.97& 536& 2.91$\times$10$^{-3}$&Y&Y\\  
 38 15 24 $-$ 37 15 23 & 340551.30& 568& 2.84$\times$10$^{-3}$&N&N\\  
 38 15 23 $-$ 37 15 22 & 340551.30& 568& 2.84$\times$10$^{-3}$&N&N\\  
 38 9 30 $-$ 37 9 29 & 340576.00& 409& 3.17$\times$10$^{-3}$&N&N\\  
 38 9 29 $-$ 37 9 28 & 340576.00& 409& 3.17$\times$10$^{-3}$&N&N\\  
 38 8 30 $-$ 37 8 29 & 340742.90& 390& 3.22$\times$10$^{-3}$&Y&Y\\  
 38 8 31 $-$ 37 8 30 & 340742.90& 390& 3.22$\times$10$^{-3}$&Y&Y\\  
 38 4 35 $-$ 37 4 34 & 340972.71& 337& 3.34$\times$10$^{-3}$&N&N\\  
 38 7 32 $-$ 37 7 31 & 341025.64& 373& 3.26$\times$10$^{-3}$&Y&Y\\  
 38 7 31 $-$ 37 7 30 & 341033.90& 373& 3.26$\times$10$^{-3}$&N&N\\  
 38 6 33 $-$ 37 6 32 & 341468.71& 359& 3.31$\times$10$^{-3}$&N&N\\  
 38 6 32 $-$ 37 6 31 & 341603.25& 359& 3.31$\times$10$^{-3}$&Y&Y\\  
 40 1 40 $-$ 39 1 39 & 341703.65& 339& 3.39$\times$10$^{-3}$&Y&Y\\  
 40 0 40 $-$ 39 0 39 & 341710.58& 339& 3.39$\times$10$^{-3}$&Y&Y\\  
 38 5 34 $-$ 37 5 33 & 341852.71& 347& 3.34$\times$10$^{-3}$&N&N\\  
 15 5 11 $-$ 14 4 10 & 342651.89& 79& 1.39$\times$10$^{-4}$&Y&Y\\  
 15 5 10 $-$ 14 4 11 & 342677.56& 79& 1.39$\times$10$^{-4}$&Y&Y\\  
 38 5 33 $-$ 37 5 32 & 343194.57& 347& 3.38$\times$10$^{-3}$&Y&N\\  
 39 3 37 $-$ 38 3 36 & 345921.20& 344& 3.50$\times$10$^{-3}$&Y&Y\\  
 38 4 34 $-$ 37 4 33 & 346874.29& 339& 3.52$\times$10$^{-3}$&N&N\\  
 38 3 35 $-$ 37 3 34 & 346983.83& 333& 3.53$\times$10$^{-3}$&N&N\\  
 39 2 37 $-$ 38 2 36 & 348260.57& 344& 3.57$\times$10$^{-3}$&Y&Y\\  
 40 2 39 $-$ 39 2 38 & 348344.53& 351& 3.58$\times$10$^{-3}$&N&N\\  
 40 1 39 $-$ 39 1 38 & 348553.31& 351& 3.59$\times$10$^{-3}$&N&N\\  
 39 12 27 $-$ 38 12 26 & 349379.90& 495& 3.29$\times$10$^{-3}$&N&N\\  
 39 12 28 $-$ 38 12 27 & 349379.90& 495& 3.29$\times$10$^{-3}$&N&N\\  
 39 11 29 $-$ 38 11 28 & 349392.34& 470& 3.34$\times$10$^{-3}$&Y&Y\\  
 39 11 28 $-$ 38 11 27 & 349392.34& 470& 3.34$\times$10$^{-3}$&Y&Y\\  
 39 13 27 $-$ 38 13 26 & 349396.20& 523& 3.23$\times$10$^{-3}$&Y&Y\\  
 39 13 26 $-$ 38 13 25 & 349396.20& 523& 3.23$\times$10$^{-3}$&Y&Y\\  
 39 14 25 $-$ 38 14 24 & 349435.13& 553& 3.17$\times$10$^{-3}$&N&N\\  
 39 14 26 $-$ 38 14 25 & 349435.13& 553& 3.17$\times$10$^{-3}$&N&N\\  
 39 10 30 $-$ 38 10 29 & 349442.94& 446& 3.40$\times$10$^{-3}$&Y&Y\\  
 39 10 29 $-$ 38 10 28 & 349442.94& 446& 3.40$\times$10$^{-3}$&Y&Y\\  
 39 9 31 $-$ 38 9 30 & 349547.01& 425& 3.44$\times$10$^{-3}$&N&N\\  
 39 9 30 $-$ 38 9 29 & 349547.03& 425& 3.44$\times$10$^{-3}$&N&N\\  
 39 8 32 $-$ 38 8 31 & 349730.78& 407& 3.49$\times$10$^{-3}$&N&N\\  
 39 8 31 $-$ 38 8 30 & 349731.30& 407& 3.49$\times$10$^{-3}$&N&N\\  
 39 4 36 $-$ 38 4 35 & 349796.03& 354& 3.61$\times$10$^{-3}$&Y&Y\\  
 39 7 33 $-$ 38 7 32 & 350039.41& 390& 3.54$\times$10$^{-3}$&N&N\\  
 39 7 32 $-$ 38 7 31 & 350050.91& 390& 3.54$\times$10$^{-3}$&N&N\\  
 41 1 41 $-$ 40 1 40 & 350139.63& 356& 3.65$\times$10$^{-3}$&Y&Y\\  
 41 0 41 $-$ 40 0 40 & 350145.09& 356& 3.65$\times$10$^{-3}$&Y&Y\\  
 39 6 34 $-$ 38 6 33 & 350511.72& 376& 3.58$\times$10$^{-3}$&Y&N\\  
 39 6 33 $-$ 38 6 32 & 350687.84& 376& 3.59$\times$10$^{-3}$&Y&Y\\  
 39 5 35 $-$ 38 5 34 & 350866.40& 364& 3.62$\times$10$^{-3}$&N&N\\  
 39 5 34 $-$ 38 5 33 & 352500.71& 364& 3.67$\times$10$^{-3}$&N&N\\  
 11 6 6 $-$ 10 5 5 & 353234.74& 68& 2.06$\times$10$^{-4}$&Y&N\\  
 11 6 5 $-$ 10 5 6 & 353234.76& 68& 2.06$\times$10$^{-4}$&Y&N\\  
 40 3 38 $-$ 39 3 37 & 354476.66& 361& 3.77$\times$10$^{-3}$&Y&Y\\  
 39 3 36 $-$ 38 3 35 & 355755.96& 350& 3.81$\times$10$^{-3}$&N&N\\  
 39 4 35 $-$ 38 4 34 & 356276.74& 356& 3.81$\times$10$^{-3}$&Y&Y\\  
 40 2 38 $-$ 39 2 37 & 356546.15& 361& 3.84$\times$10$^{-3}$&Y&Y\\  
 41 2 40 $-$ 40 2 39 & 356789.59& 368& 3.85$\times$10$^{-3}$&Y&N\\  
 41 1 40 $-$ 40 1 39 & 356960.42& 368& 3.86$\times$10$^{-3}$&N&N\\  
 40 12 29 $-$ 39 12 28 & 358326.20& 512& 3.57$\times$10$^{-3}$&N&N\\  
 40 12 28 $-$ 39 12 27 & 358326.20& 512& 3.57$\times$10$^{-3}$&N&N\\  
 40 13 27 $-$ 39 13 26 & 358339.01& 540& 3.51$\times$10$^{-3}$&N&N\\  
 40 13 28 $-$ 39 13 27 & 358339.01& 540& 3.51$\times$10$^{-3}$&N&N\\  
 40 11 29 $-$ 39 11 28 & 358344.02& 487& 3.62$\times$10$^{-3}$&Y&Y\\  
 40 11 30 $-$ 39 11 29 & 358344.02& 487& 3.62$\times$10$^{-3}$&Y&Y\\  
 40 14 27 $-$ 39 14 26 & 358375.84& 570& 3.44$\times$10$^{-3}$&N&N\\  
 40 14 26 $-$ 39 14 25 & 358375.84& 570& 3.44$\times$10$^{-3}$&N&N\\  
 40 10 31 $-$ 39 10 30 & 358402.65& 464& 3.68$\times$10$^{-3}$&Y&N\\  
 40 10 30 $-$ 39 10 29 & 358402.65& 464& 3.68$\times$10$^{-3}$&Y&N\\  
 40 9 32 $-$ 39 9 31 & 358518.65& 443& 3.73$\times$10$^{-3}$&N&N\\  
 40 9 31 $-$ 39 9 30 & 358518.67& 443& 3.73$\times$10$^{-3}$&N&N\\  
 42 1 42 $-$ 41 1 41 & 358573.80& 373& 3.92$\times$10$^{-3}$&Y&Y\\  
 42 0 42 $-$ 41 0 41 & 358578.09& 373& 3.92$\times$10$^{-3}$&Y&Y\\  
 40 4 37 $-$ 39 4 36 & 358595.53& 371& 3.89$\times$10$^{-3}$&Y&Y\\  
 40 8 33 $-$ 39 8 32 & 358720.46& 424& 3.78$\times$10$^{-3}$&Y&Y\\  
 40 8 32 $-$ 39 8 31 & 358721.21& 424& 3.78$\times$10$^{-3}$&Y&Y\\  
 40 7 34 $-$ 39 7 33 & 359056.12& 407& 3.82$\times$10$^{-3}$&Y&Y\\  
 40 7 33 $-$ 39 7 32 & 359071.92& 407& 3.82$\times$10$^{-3}$&N&N\\  
 40 6 35 $-$ 39 6 34 & 359556.50& 393& 3.87$\times$10$^{-3}$&Y&Y\\  
 40 6 34 $-$ 39 6 33 & 359785.04& 393& 3.88$\times$10$^{-3}$&Y&N\\  
 40 5 36 $-$ 39 5 35 & 359870.11& 381& 3.91$\times$10$^{-3}$&Y&Y\\  
 17 5 12 $-$ 16 4 13 & 360461.27& 94& 1.52$\times$10$^{-4}$&Y&Y\\  
 40 5 35 $-$ 39 5 34 & 361840.75& 382& 3.97$\times$10$^{-3}$&N&N\\  
 12 6 7 $-$ 11 5 6 & 362187.71& 74& 2.10$\times$10$^{-4}$&Y&Y\\  
 12 6 6 $-$ 11 5 7 & 362187.75& 74& 2.10$\times$10$^{-4}$&Y&Y\\  
\hline

\end{longtable} 
\begin{table*}
\caption{Detected lines of C$_2$H$_3$CN \label{tab.lines}}
\begin{tabular}{c c c c c }
\hline
\hline
 Transitions & Frequency & $E_\mathrm{up}$ & $A_\mathrm{ij}$ & Blended B$^a$ \\
             & (MHz) & (K) & (s$^{-1}$) &  \\
\hline	
 36 1 36 $-$ 35 1 35 & 329191.68& 296& 2.98$\times$10$^{-3}$&N\\  
 36 0 36 $-$ 35 0 35 & 329462.36& 296& 2.98$\times$10$^{-3}$&Y\\  
 35 1 34 $-$ 34 1 33 & 331086.67& 292& 3.02$\times$10$^{-3}$&N\\  
 35 3 33 $-$ 34 3 32 & 332166.70& 307& 3.04$\times$10$^{-3}$&Y\\  
 35 5 31 $-$ 34 5 30 & 332775.70& 341& 3.01$\times$10$^{-3}$&N\\  
 35 4 32 $-$ 34 4 31 & 333047.30& 322& 3.04$\times$10$^{-3}$&Y\\  
 35 4 31 $-$ 34 4 30 & 333764.56& 322& 3.06$\times$10$^{-3}$&Y\\  
 36 2 35 $-$ 35 2 34 & 337039.73& 310& 3.19$\times$10$^{-3}$&Y\\  
 35 3 32 $-$ 34 3 31 & 337050.90& 308& 3.17$\times$10$^{-3}$&Y\\  
 37 1 37 $-$ 36 1 36 & 338213.51& 312& 3.23$\times$10$^{-3}$&N\\  
 35 2 33 $-$ 34 2 32 & 338278.15& 300& 3.22$\times$10$^{-3}$&N\\  
 37 0 37 $-$ 36 0 36 & 338447.69& 312& 3.24$\times$10$^{-3}$&N\\  
 36 1 35 $-$ 35 1 34 & 340047.92& 308& 3.28$\times$10$^{-3}$&Y\\  
 36 3 34 $-$ 35 3 33 & 341563.75& 323& 3.30$\times$10$^{-3}$&N\\  
 36 6 31 $-$ 35 6 30 & 342052.94& 381& 3.25$\times$10$^{-3}$&N\\  
 36 6 30 $-$ 35 6 29 & 342055.21& 381& 3.25$\times$10$^{-3}$&Y\\  
 36 5 32 $-$ 35 5 31 & 342317.55& 358& 3.29$\times$10$^{-3}$&Y\\  
 36 5 31 $-$ 35 5 30 & 342375.56& 358& 3.29$\times$10$^{-3}$&Y\\  
 36 4 33 $-$ 35 4 32 & 342585.47& 338& 3.32$\times$10$^{-3}$&N\\  
 36 4 32 $-$ 35 4 31 & 343446.53& 339& 3.34$\times$10$^{-3}$&Y\\  
 37 2 36 $-$ 36 2 35 & 346184.94& 326& 3.45$\times$10$^{-3}$&Y\\  
 36 3 33 $-$ 35 3 32 & 346943.07& 325& 3.46$\times$10$^{-3}$&Y\\  
 38 1 38 $-$ 37 1 37 & 347232.01& 329& 3.50$\times$10$^{-3}$&N\\  
 38 0 38 $-$ 37 0 37 & 347434.23& 329& 3.50$\times$10$^{-3}$&N\\  
 36 2 34 $-$ 35 2 33 & 347759.03& 317& 3.50$\times$10$^{-3}$&N\\  
 37 1 36 $-$ 36 1 35 & 348991.44& 325& 3.54$\times$10$^{-3}$&Y\\  
 37 3 35 $-$ 36 3 34 & 350947.22& 340& 3.59$\times$10$^{-3}$&Y\\  
 37 5 33 $-$ 36 5 32 & 351861.48& 375& 3.57$\times$10$^{-3}$&N\\  
 37 5 32 $-$ 36 5 31 & 351935.23& 375& 3.58$\times$10$^{-3}$&N\\  
 37 4 34 $-$ 36 4 33 & 352120.64& 355& 3.61$\times$10$^{-3}$&Y\\  
 37 4 33 $-$ 36 4 32 & 353147.23& 356& 3.64$\times$10$^{-3}$&N\\  
 38 2 37 $-$ 37 2 36 & 355317.96& 343& 3.74$\times$10$^{-3}$&N\\  
 39 1 39 $-$ 38 1 38 & 356247.42& 346& 3.78$\times$10$^{-3}$&Y\\  
 39 0 39 $-$ 38 0 38 & 356421.72& 346& 3.78$\times$10$^{-3}$&Y\\  
 37 3 34 $-$ 36 3 33 & 356832.00& 342& 3.77$\times$10$^{-3}$&Y\\  
 37 2 35 $-$ 36 2 34 & 357202.09& 334& 3.80$\times$10$^{-3}$&Y\\  
 38 1 37 $-$ 37 1 36 & 357920.91& 342& 3.82$\times$10$^{-3}$&Y\\  
 38 3 36 $-$ 37 3 35 & 360316.58& 357& 3.89$\times$10$^{-3}$&N\\  
 38 6 33 $-$ 37 6 32 & 361097.14& 416& 3.84$\times$10$^{-3}$&N\\  
 38 6 32 $-$ 37 6 31 & 361101.24& 416& 3.84$\times$10$^{-3}$&Y\\  
 38 5 34 $-$ 37 5 33 & 361407.35& 392& 3.88$\times$10$^{-3}$&Y\\  
 38 5 33 $-$ 37 5 32 & 361500.42& 392& 3.88$\times$10$^{-3}$&Y\\  
 38 4 35 $-$ 37 4 34 & 361652.04& 373& 3.91$\times$10$^{-3}$&Y\\  

\hline
\end{tabular}
\tablefoot{$^a$Y means line is blended with another molecule.}
\clearpage
\end{table*} 
\begin{table*}
\caption{Detected lines of H$C_3$N \label{tab.lines}}
\begin{tabular}{c c c c c c }
\hline
\hline
Transitions & Frequency & $E_\mathrm{up}$ & $A_\mathrm{ij}$ & Blended A$^a$& Blended B$^a$ \\
            & (MHz) & (K) & (s$^{-1}$) & & \\
\hline	
 37 $-$ 36 & 336520.08& 307& 3.05$\times$10$^{-3}$&Y&N\\
 38 $-$ 37 & 345609.01& 323& 3.30$\times$10$^{-3}$&Y&Y\\
 39 $-$ 38 & 354697.00& 341& 3.57$\times$10$^{-3}$&Y&N\\
\hline
\end{tabular}
\tablefoot{$^a$Y means line is blended with another molecule.}

\end{table*} 
\clearpage
\section{Velocity-corrected INtegrated Emission (VINE) maps}
\label{sec:appen2}
Velocity-corrected integrated emission (VINE) maps of IRAS 16293A. 

 \begin{figure*}[h] 
 \begin{center} 
\includegraphics[width=20cm, angle=0, clip =true, trim = 1cm 1.1cm 2cm 2cm]{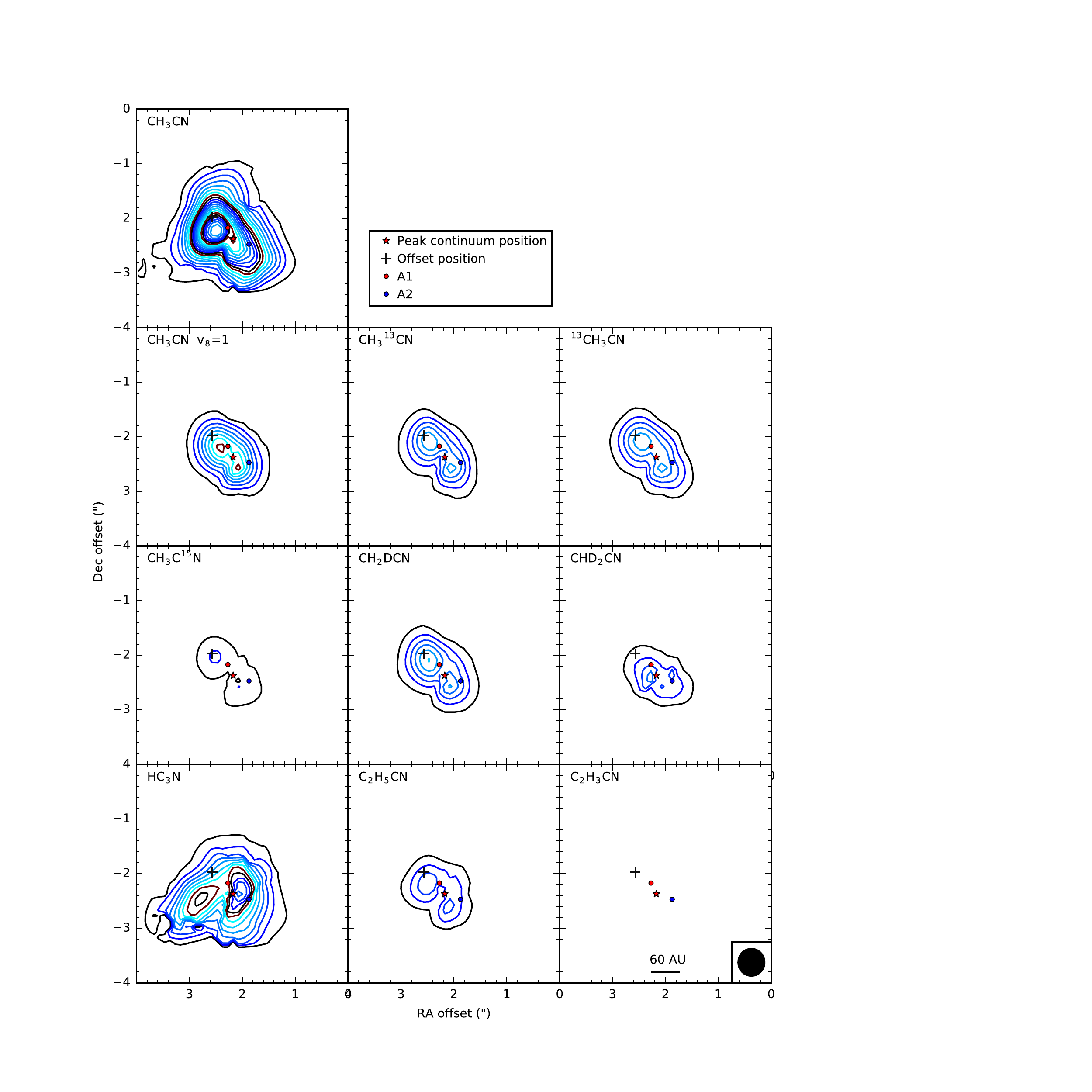}
  \end{center} 
  \vspace{-0.5cm}
 \caption{Velocity-corrected integrated emission (VINE) maps of IRAS 16293A for the 349.346\,GHz line of CH$_3$CN v=0, the 350.168\,GHz line of CH$_3$CN v$_8$=1, the 339.366\,GHz line of $^{13}$CH$_3$CN, the 349.254\,GHz of CH$_3^{13}$CN, the 338.940\,GHz of CH$_3$C$^{15}$N, the 347.110\,GHz of CH$_2$DCN, the 362.520\,GHz of CHD$_2$CN, the 329.235\,GHz of C$_2$H$_5$CN, the 341.564\,GHz of C$_2$H$_3$CN and the 336.520\,GHz of HC$_3$N. The axis show the position offset from phase centre of the observations. Contour levels start at 80\,mJy\,km\,s$^{-1}$ and increase in steps of 11\,mJy\,km\,s$^{-1}$. The red star marks the peak continuum position in the PILS dataset of IRAS 16293A . The black cross marks the offset position where the spectra analysed in this work are extracted from. The red and blue circles mark the positions of the two continuum peaks A1 and A2 respectively, found by previous authors, using coordinates that have been corrected by the rate of position angle change with year determined by \citet{Pech2010}.\label{fig:nmap_zoom}} 
 \end{figure*} 

\end{document}